\documentclass[onecolumn]{elsart1p}
\usepackage[numbers,sort&compress]{natbib}
\usepackage{listings}
\usepackage{hyperref}
\usepackage[pdftex]{graphicx}
\usepackage{color}
\usepackage{url}
\usepackage{amsmath}
\usepackage{amssymb}
\usepackage{wasysym}
\usepackage{stmaryrd}
\usepackage{rotating}
\usepackage{semantic}
\usepackage{boxedminipage}
\usepackage{fancybox}
\newlength{\sidebarwidth}
\usepackage{array}
\usepackage{times}

\definecolor{lightgray}{gray}{0.93}

\def\Cpp{{C\nolinebreak[4]\hspace{-.05em}\raisebox{.4ex}{\tiny\bf ++}}}
\def\FG{{F$^\mathrm{G}$}}
\def\G{{$\mathcal{G}$}}

\lstdefinestyle{G}{showstringspaces=false,columns=fullflexible,
  language=C++,mathescape=true,xleftmargin=1pc,
  moredelim=**[is][\color{red}]{^}{^},
  keepspaces,
  lineskip=-1pt,
  aboveskip=4pt,belowskip=3pt,
  basicstyle=\ttfamily,
  commentstyle=\rmfamily\it,
  morekeywords={fun,where,concept,model,require,refines,type,let,module,scope,import},
  keywordstyle=\color{blue}\ttfamily}
\lstdefinestyle{Java}{showstringspaces=false,columns=fullflexible,language=Java,xleftmargin=1pc,
basicstyle=\small\ttfamily,
keywordstyle=\small\ttfamily,
keepspaces,
aboveskip=4pt,belowskip=4pt}

\lstdefinestyle{ML}{showstringspaces=false,columns=fullflexible,language=ML,xleftmargin=1pc,
basicstyle=\small\ttfamily,
keywordstyle=\small\ttfamily,
keepspaces,
aboveskip=4pt,belowskip=4pt}

\lstdefinestyle{Cpp}{showstringspaces=false,columns=fullflexible,language=C++,xleftmargin=1pc,
basicstyle=\small\ttfamily,
keywordstyle=\small\ttfamily,
moredelim=**[is][\color{blue}]{^}{^},
keepspaces,
aboveskip=4pt,belowskip=4pt}

\lstdefinestyle{Scheme}{showstringspaces=false,columns=fullflexible,language=Lisp,xleftmargin=1pc,
basicstyle=\small\ttfamily,
mathescape=true,
keywordstyle=\small\ttfamily,
commentstyle=\small\ttfamily,
keepspaces,
moredelim=**[is][\color{green}]{^}{^},
moredelim=**[is][\color{blue}]{`}{`},
aboveskip=5pt,belowskip=5pt,
literate={.*}{{$\times$}}1 {lambda}{{$\lambda$}}1 {generic}{{$\Lambda$}}1 {forall}{{$\forall$}}1 {->}{{$\rightarrow$}}1 {-->}{{$\longrightarrow$}}1 
}

\lstdefinestyle{Haskell}{showstringspaces=false,columns=fullflexible,language=Haskell,xleftmargin=1pc,
basicstyle=\small\ttfamily,
keywordstyle=\small\ttfamily,
keepspaces,
aboveskip=4pt,belowskip=4pt}

\lstdefinestyle{Ada}{showstringspaces=false,columns=fullflexible,language=Ada,xleftmargin=1pc,
basicstyle=\small\ttfamily,
keywordstyle=\small\ttfamily,
keepspaces,
aboveskip=4pt,belowskip=4pt}

\newcommand{\code}[1]{\lstinline{#1}}

\newcommand{\Keyword}[1]{\emph{#1}}
\newcommand{\concept}[1]{\emph{#1}}
\newcommand{\fundef}[0]{\mathit{fundef}}
\newcommand{\funsig}[0]{\mathit{funsig}}
\newcommand{\expr}[0]{\mathit{expr}}
\newcommand{\stmt}[0]{\mathit{stmt}}
\newcommand{\decl}[0]{\mathit{decl}}
\newcommand{\type}[0]{\mathit{type}}
\newcommand{\id}[0]{\mathit{id}}
\newcommand{\tyid}[0]{\mathit{tyid}}
\newcommand{\cmem}[0]{\mathit{cmem}}
\newcommand{\cid}[0]{\mathit{cid}}
\newcommand{\clid}[0]{\mathit{clid}}
\newcommand{\assoc}[0]{\mathit{assoc}}
\newcommand{\constraint}[0]{\mathit{constraint}}
\newcommand{\pass}[0]{\mathit{pass}}
\newcommand{\polyhdr}[0]{\mathit{polyhdr}}
\newcommand{\classmem}[0]{\mathit{classmem}}
\newcommand{\mem}[0]{\mathit{mem}}

\newcommand{\moduleid}[0]{\mathit{mid}}
\newcommand{\cpt}[1]{{\emph{#1}}}
\newcommand{\MLF}[0]{\ensuremath{\mathrm{ML}^\mathrm{F}}}

\newcommand{\where}[0]{\texttt{where}}

\newcommand{\fun}[0]{\texttt{fun}}
\newcommand{\arrow}[0]{\texttt{->}}

\newcommand{\LT}[0]{\texttt{<}}
\newcommand{\GT}[0]{\texttt{>}}
\newcommand{\LP}[0]{\texttt{(}}
\newcommand{\RP}[0]{\texttt{)}}
\newcommand{\LB}[0]{\texttt{\{}}
\newcommand{\RB}[0]{\texttt{\}}}
\newcommand{\rep}[1]{\overline{#1}}

\def\dstart{\hbox to \hsize{\vrule depth 4pt\hrulefill\vrule depth 4pt}}
\def\dend{\hbox to \hsize{\vrule height 4pt\hrulefill\vrule height 4pt}}

\newenvironment{fdisplay}[2]{\begin{figure}[tbp]\caption{#1}\label{#2}\vbox\bgroup\bigskip\vspace{-5pt}\dstart}{\vspace{-5pt}\dend\bigskip\egroup\end{figure}}

\begin{document}

\begin{frontmatter}

\title{A Language for Generic Programming in the Large}
\author[1]{Jeremy G. Siek}\ead{jeremy.siek@colorado.edu}
and \author[2]{Andrew Lumsdaine}\ead{lums@osl.iu.edu}
\address[1]{Deptartment of Electrical and Computer Engineering, University of Colorado at Boulder, USA}
\address[2]{Computer Science Department, Indiana University, USA}

\begin{abstract}
Generic programming is an effective methodology for developing
reusable software libraries.
Many programming languages provide generics and have features for
describing interfaces, but none completely support the idioms used in
generic programming.
To address this need we developed the language \G{}.
The central feature of \G{} is the \code{concept}, a mechanism for
organizing constraints on generics that is inspired by the needs of
modern \Cpp{} libraries. 
\G{} provides modular type checking and separate compilation (even of
generics).
These characteristics support modular software development, especially
the smooth integration of independently developed components.
In this article we present the rationale for the design of \G{} and
demonstrate the expressiveness of \G{} with two case studies: porting
the Standard Template Library and the Boost Graph Library from \Cpp{}
to \G{}. The design of \G{} shares much in common with the
\code{concept} extension proposed for the next \Cpp{} Standard (the
authors participated in its design) but there are important
differences described in this article.
\end{abstract}

\begin{keyword}
programming language design \sep generic programming \sep generics \sep polymorphism \sep concepts \sep associated types \sep software reuse \sep type classes \sep modules \sep signatures \sep functors
\sep virtual types
\end{keyword}

\end{frontmatter}

\maketitle

\section{Introduction}
\label{sec:introduction}

The 1968 NATO Conference on Software Engineering identified a software
crisis affecting large systems such as IBM's OS/360 and the SABRE
airline reservation
system~\cite{randell79:_software_in_1968,frederick78:_mythic_man_month}.
At this conference McIlroy gave an invited talk entitled
\emph{Mass-produced Software Components}~\cite{mcilroy69:_components}
proposing the systematic creation of reusable software components as a
solution to the software crisis.  He observed that most software is
created from similar building blocks, so programmer productivity would
be increased if a standard set of blocks could be shared among many
software products.
We are beginning to see the benefits of software reuse; Douglas
McIlroy's vision is gradually becoming a reality. The number of
commercial and open source software libraries is steadily growing and
application builders often turn to libraries for user-interface
components, database access, report creation, numerical routines, and
network communication, to name a few. Furthermore, larger software
companies have benefited from the creation of in-house domain-specific
libraries which they use to support entire software product
lines~\cite{clements02:_prod_lines}.

As the field of software engineering progresses, we learn better
techniques for building reusable software.  In the 1980s Musser and
Stepanov developed a methodology for creating highly reusable
algorithm
libraries~\cite{KMS81,Musser:1989:GP,Musser87,Kershenbaum88}, using
the term \emph{generic programming} for their work.\footnote{The term
  generic programming is often used to mean any use of generics, i.e.,
  any use of parametric polymorphism or templates.  The term is also
  used in the functional programming community for function generation
  based on algebraic datatypes, i.e., polytypic programming. Here, we
  use generic programming solely in the sense of Musser and Stepanov.}
Their approach was novel in that they wrote algorithms not in terms of
particular data structures but rather in terms of abstract
requirements on structures based on the needs of the algorithm.
Such generic algorithms could operate on any data structure provided
that it meet the specified requirements.
Preliminary versions of their generic algorithms were implemented in
Scheme, Ada, and C.
In the early 1990s Stepanov and Musser took advantage of the template
feature of \Cpp{}~\cite{stroustrup88:_param_types} to construct the
Standard Template Library
(STL)~\cite{stepa.lee-1994:the.s:TR,austern99:_gener_progr_stl}.
The STL became part of the \Cpp{} Standard, which brought generic
programming into the mainstream.  Since then, the methodology has been
successfully applied to the creation of libraries in numerous
domains~\cite{koethe99:_reusable_vision,siek02:_bgl,boissonnat99:_cgal,pitt01:_bioinf_template_lib,troyer:_alps}.

The ease with which programmers develop and use generic libraries
varies greatly depending on the language features available for
expressing polymorphism and requirements on type parameters. 
In 2003 we performed a comparative study of modern language support
for generic programming~\cite{comparing_generic_programming03}.  The
initial study included \Cpp{}, SML, Haskel, Eiffel, Java, and C\#, and
we evaluated the languages by porting a representative subset of the
Boost Graph Library~\cite{siek02:_bgl} to each of them.  We recently
updated the study to include OCaml and Cecil~\cite{Garcia:2007fk}.
While some languages performed quite well, none were ideal for generic
programming.

Unsatisfied with the state of the art, we began to investigate how to
improve language support for generic programming. In general we wanted
a language that could express the idioms of generic programming while
also providing \emph{modular type checking} and \emph{separate
  compilation}. In the context of generics, modular type checking
means that a generic function or class can be type checked
independently of any instantiation and that the type check
guarantees that any well-typed instantiation will produce well-typed
code.  Separate compilation is the ability to compile a generic
function to native assembly code that can be linked into an
application in constant time.

Our desire for modular type checking was a reaction to serious
problems that plague the development and use of \Cpp{} template
libraries. A \Cpp{} template definition is not type checked until
after it is instantiated, making templates difficult to validate in
isolation.  Even worse, clients of template libraries are exposed to
confusing error messages when they accidentally misuse the library.
For example, the following code tries to use \code{stable_sort} with
the iterators from the \code{list} class.

\begin{lstlisting}[language=C++]
std::list<int> l;
std::stable_sort(l.begin(), l.end());
\end{lstlisting}

\noindent Fig.~\ref{fig:stable-sort-error} shows a portion  
of the error message from GNU \Cpp{}. The error message includes
functions and types that the client should not have to know about such
as \code{__inplace_stable_sort} and \code{_List_iterator}.  It is not
clear from the error message who is responsible for the error.  The
error message points inside the STL so the client might conclude that
there is an error in the STL.  This problem is not specific to the GNU
\Cpp{} implementation, but is instead a symptom of the delayed type
checking mandated by the \Cpp{} language definition.

\begin{figure*}[htbp]
  \centering
\begin{lstlisting}[basicstyle=\scriptsize,keywordstyle=\scriptsize,lineskip=-2pt]
stl_algo.h: In function `void std::__inplace_stable_sort(_RandomAccessIter, _RandomAccessIter) 
    [with _RandomAccessIter = std::_List_iterator<int, int&, int*>]':
stl_algo.h:2565:   instantiated from `void std::stable_sort(_RandomAccessIter, _RandomAccessIter) 
    [with _RandomAccessIter = std::_List_iterator<int, int&, int*>]'
stable_sort_error.cpp:5:   instantiated from here
stl_algo.h:2345: error: no match for `std::_List_iterator<int, int&, int*>& - std::_List_iterator<int, int&, int*>&' operator
stl_algo.h:2565:   instantiated from `void std::stable_sort(_RandomAccessIter, _RandomAccessIter) 
    [with _RandomAccessIter = std::_List_iterator<int, int&, int*>]'
stable_sort_error.cpp:5:   instantiated from here
stl_algo.h:2349: error: no match for `std::_List_iterator<int, int&, int*>& - std::_List_iterator<int, int&, int*>&' operator
stl_algo.h:2352: error: no match for `std::_List_iterator<int, int&, int*>& - std::_List_iterator<int, int&, int*>&' operator
\end{lstlisting}
  \caption{A portion of the error message from a misuse of \code{stable_sort}.}
  \label{fig:stable-sort-error}
\end{figure*}

Our desire for separate compilation was driven by the increasingly
long compile times we (and others) were experiencing when composing
sophisticated template libraries. With \Cpp{} templates, the
compilation time of an application is a function of the amount of code
in the application plus the amount of code in all template libraries
used by the application (both directly and indirectly). We would much
prefer a scenario where generic libraries can be separately compiled
so that the compilation time of an application is just a function of
the amount of code in the application.

With these desiderata in hand we began laying the theoretical
groundwork by developing the calculus \FG{}~\cite{Siek:2005mf}.  \FG{}
is based on System F~\cite{GIRARD72,REYNOLDS83}, the standard calculus
for parametric polymorphism, and like System F, \FG{} has a modular
type checker and can be separately compiled. In addition, \FG{}
provides features for precisely expressing constraints on generics,
introducing the \code{concept} feature with support for associated
types and same-type constraints. The main technical challenge overcome
in \FG{} is dealing with type equality inside of generic functions.
One of the key design choices in \FG{} is that models are lexically
scoped, making \FG{} more modular than Haskell in this regard. (We
discuss this in more detail in Section~\ref{sec:scoped-models}.)
Concurrently with our work on \FG{}, Chakravarty, Keller, and Peyton
Jones responded to our comparative study by developing an extension to
Haskell to support associated
types~\cite{chakravarty04:_assoc_types,chakravarty05:_assoc_type_syn}.

The next step after \FG{} was to add two more features needed to
express generic libraries: concept-based overloading (used for
algorithm specialization) and implicit argument deduction.  Fully
general implicit argument deduction is non-trivial in the presence of
first-class polymophism (which is present in \G{}), but some mild
restrictions make the problem tractable
(Section~\ref{sec:implicit-instantiation}). However, we discovered a a
deep tension between concept-based overloading and separate
compilation~\cite{jaakko06:_algo_spec}.
At this point our work bifurcated into two language designs: the
language \G{} which supports separate compilation and only a basic
form of concept-based overloading~\cite{Siek:2005lr,siek05:_g_stl},
and the concepts extension to \Cpp{}~\cite{gregor06:_concepts}, which
provides full support for concept-based overloading but not separate
compilation.
For the next revision of the \Cpp{} Standard, popularly referred to as
\Cpp{}0X, separate compilation for templates was not practical because
the language already included template specialization, a feature that
is also deeply incompatible with separate compilation.  Thus, for
\Cpp{}0X it made sense to provide full support for concept-based
overloading. For \G{} we placed separate compilation as a higher
priority, leaving out template specialization and requiring
programmers to work around the lack of full concept-based overloading
(see Section X).

Table~\ref{generics_feature_comparison} shows the results of our
comparative study of language support for generic
programming~\cite{Garcia:2007fk} augmented with new columns for
\Cpp{}0X and \G{} and augmented with three new rows: modular type
checking (previously part of ``separate compilation''), lexically
scoped models, and concept-based overloading. Table~\ref{tab:glossary}
gives a brief description of the evaluation criteria.

\newcolumntype{C}{>{\hfill\arraybackslash}p{.4in}<{\hfill\hfill}}

\newcommand{\featY}{\CIRCLE} 
\newcommand{\featP}{\begin{sideways}\LEFTcircle\end{sideways}} 
\newcommand{\featN}{\Circle} 

\newcommand{\featA}{\hspace{4.25pt}\featY$^{*}$} 
\newcommand{\featB}{\hspace{4.25pt}\featY$^{\dagger}$} 

\begin{table*}[t]
\center
\caption{The level of support for generic programming in several languages.
A black circle indicates full support for the feature or characteristic whereas 
a white circle indices lack of support. 
The rating of ``-'' in the \protect\Cpp{} column indicates that while
\protect\Cpp{} does not explicitly support the feature, one can still program
as if the feature were supported.
\label{generics_feature_comparison}}
\vspace{5pt}
\begin{tabular}{rCCCCCCC|CC}
\hline \hline
{} & \Cpp{} & SML & OCaml & Haskell & Java & C\# & Cecil & \Cpp{}0X & \G{} \\
\hline
Multi-type concepts    & -      & \featY & \featN & \featA & \featN & \featN & \featP & \featY & \featY \\
Multiple constraints   & -      & \featP & \featP & \featY & \featY & \featY & \featY & \featY & \featY\\
Associated type access & \featY & \featY & \featP & \featB & \featP & \featP & \featP & \featY & \featY\\
Constraints on assoc. types & - & \featY & \featY & \featB & \featP & \featP & \featY & \featY & \featY\\
Retroactive modeling   & -      & \featY & \featY & \featY & \featN & \featN & \featY & \featY & \featY\\
Type aliases           & \featY & \featY & \featY & \featY & \featN & \featN & \featN & \featY & \featY\\
Separate compilation   & \featN & \featY & \featY & \featY & \featY & \featY & \featP & \featN & \featY\\
Implicit arg. deduction & \featY & \featN & \featY & \featY & \featY & \featY & \featP & \featY & \featY\\ \hline\hline
Modular type checking   &  \featN&  \featY & \featP & \featY& \featY &  \featY & \featP & \featP& \featY \\
Lexically scoped models & \featN & \featY & \featN & \featN & \featN & \featN & \featN & \featN & \featY \\
Concept-based overloading & \featY & \featN & \featN & \featN & \featN & \featN & \featY & \featY & \featP\\
\hline \hline
\end{tabular} \\
\raggedright {\small
$^{*}$Using the multi-parameter type class extension to Haskell~\cite{jones97type}.\\
$^{\dagger}$Using the proposed associated types extension to Haskell~\cite{chakravarty05:_assoc_type_syn}.
}
\end{table*}

\begin{table*}
  \centering
  \caption{Glossary of Evaluation Criteria\label{tab:glossary}}
\vspace{5pt}
  \begin{tabular}{lp{3.8in}}
\hline\hline
Criterion & Definition \\
\hline
Multi-type concepts & Multiple types can be simultaneously constrained. \\[-0.5ex]
Multiple constraints &  More than one constraint can be
placed on a type parameter.  \\[-0.5ex]
Associated type access & Types can be
mapped to other types within the context of a generic function.\\[-0.5ex]
Constraints on associated types &  Concepts
may include constraints on associated types.\\[-0.5ex]
Retroactive modeling & The ability to
  add new modeling relationships after a type has been defined.\\[-0.5ex]
Type aliases &  A mechanism for
  creating shorter names for types is provided.\\[-0.5ex]
Separate compilation &  Generic
  functions can be compiled independently of calls to them.\\[-0.5ex]
Implicit argument deduction & The arguments for the
type parameters of a generic function can be deduced and do not
need to be explicitly provided by the programmer.\\[-0.5ex]
Modular type checking &  Generic
  functions can be compiled independently of calls to them.\\[-0.5ex]
Lexically scoped models & Model declarations are treated like any
  other declaration, and are in scope for the remainder of
  enclosing namespace. Models may be explicitly imported
  from other namespaces. \\[-0.5ex]
Concept-based overloading & There can be multiple generic
  functions with the same name but differing constraints.
  For a particular call, the most specific overload is chosen.\\
\hline\hline
  \end{tabular}  
\end{table*}

The rest of this article describes the design of \G{} in detail.  We
review the essential ideas of generic programming and survey of the
idioms used in the Standard Template Library
(Section~\ref{sec:generic-programming}). This provides the motivation
for the design of the language features in \G{}
(Section~\ref{sec:overview-g}). We then evaluate \G{} with respect to
a port of the Standard Template Library
(Section~\ref{sec:stl-implementation}) and the Boost Graph Library
(Section~\ref{sec:bgl-implementation}).  We conclude with a survey of
related work (Section~\ref{sec:related-work}) and with the future
directions for our work (Section~\ref{sec:conclusion}).

This article is an updated and greatly extended version of
\cite{siek05:_g_stl}, providing a more detailed rationale for the
design of \G{} and extending our previous comparative study to include
\G{} by evaluating a port of the Boost Graph Library to \G{}.

\section{Generic Programming and the STL}
\label{sec:generic-programming}

Fig.~\ref{def:gp} reproduces the standard definition of generic
programming from Jazayeri, Musser, and
Loos~\cite{jazayeri98:_generic_programming}. 
The generic programming methodology always consists of the following
steps: 1) identify a family of useful and efficient concrete
algorithms with some commonality,
2) resolve the differences by forming higher-level abstractions,
and 3) lift the concrete algorithms so they operate on these new
abstractions.
When applicable, there is a fourth step to implement automatic
selection of the best algorithm, as described in Fig.~\ref{def:gp}.

\begin{figure}[tbp]
\newlength{\pagewidth}
\setlength{\pagewidth}{\columnwidth}
\addtolength{\pagewidth}{-2\fboxsep}
\addtolength{\pagewidth}{-2\fboxrule}
\addtolength{\pagewidth}{-1em}
\setlength{\fboxsep}{12pt} 
\begin{boxedminipage}{\pagewidth}
{\small
Generic programming is a sub-discipline of computer science that deals
with finding abstract representations of efficient algorithms, data
structures, and other software concepts, and with their systematic
organization. The goal of generic programming is to express algorithms
and data structures in a broadly adaptable, interoperable form that
allows their direct use in software construction. Key ideas include:
\begin{itemize}
\item Expressing algorithms with minimal assumptions about data
      abstractions, and vice versa, thus making them as interoperable
      as possible.
\item Lifting of a concrete algorithm to as general a level as
      possible without losing efficiency; i.e., the most abstract form
      such that when specialized back to the concrete case the result
      is just as efficient as the original algorithm.
\item When the result of lifting is not general enough to cover all
      uses of an algorithm, additionally providing a more general
      form, but ensuring that the most efficient specialized form is
      automatically chosen when applicable.
\item Providing more than one generic algorithm for the same purpose and
      at the same level of abstraction, when none dominates the others
      in efficiency for all inputs. This introduces the necessity to
      provide sufficiently precise characterizations of the domain for
      which each algorithm is the most efficient. 
\end{itemize}
}
\end{boxedminipage}
\caption{Definition of Generic Programming from Jazayeri, Musser, and Loos\cite{jazayeri98:_generic_programming}}\label{def:gp}
\end{figure}

\subsection{Type requirements, concepts, and models}

The \code{merge} algorithm from the STL, shown in
Fig.~\ref{fig:merge}, serves as a good example of generic programming.
The algorithm does not directly work on a particular data structure,
such as an array or linked list, but instead operates on an abstract
entity, a concept. A \Keyword{concept} is a collection of requirements
on a type, or to look at it a different way, it is the set of all
types that satisfy the requirements. For example, the \concept{Input
  Iterator} concept requires that the type have an increment and
dereference operation, and that both are constant-time operations.
(We italicize concept names.)  A type that meets the requirements is
said to \Keyword{model} the concept.  (It helps to read ``models'' as
``implements''.)  For example, the models of the \concept{Input
  Iterator} concept include the built-in pointer types, such as
\texttt{int*}, the iterator type for the \code{std::list} class, and
the \code{istream_iterator} adaptor.
Constraints on type parameters are primarily expressed by requiring
the corresponding type arguments to model certain concepts.  In the
\code{merge} template, the argument for \texttt{InIter1} is required
to model the \concept{Input Iterator} concept.  Type requirements are
not expressible in \Cpp{}, so the convention is to specify type
requirements in comments or documentation as in Fig.~\ref{fig:merge}.

\begin{fdisplay}{The \code{merge} algorithm in C++.}{fig:merge}
\begin{lstlisting}
template<typename InIter1, typename InIter2, typename OutIter>
  // where InIter1 models Input Iterator, InIter2 models Input Iterator.
  //     OutIter models Output Iterator, writing the value_type of InIter1.
  //     The value_type of InIter1 and InIter2 are the same type.
  //     The value_type of InIter1 is Less Than Comparable.
OutIter merge(InIter1 first1, InIter1 last1,
              InIter2 first2, InIter2 last2, OutIter result) {
  while (first1 != last1 && first2 != last2) {
    if (*first2 < *first1) {
      *result = *first2; ++first2;
    } else {
      *result = *first1; ++first1;
    }
    ++result;
  }
  return copy(first2, last2, copy(first1, last1, result));
}
\end{lstlisting}
\end{fdisplay}

The type requirements for \code{merge} refer to relationships between
types, such as the \code{value_type} of \code{InIter1}.  This is an
example of an \Keyword{associated type}, which maps between types that
are part of a concept.  The \code{merge} algorithm also needs to
express that the \code{value_type} of \code{InIter1} and
\code{InIter2} are the same, which we call \Keyword{same-type
  constraints}.  Furthermore, the \code{merge} algorithm includes an
example of how associated types and modeling requirements can be
combined: the \code{value_type} of the input iterators is required to
be \concept{Less Than Comparable}.

Fig.~\ref{fig:InputIterator} shows the definition of the \cpt{Input
  Iterator} concept following the presentation style used in the SGI
STL documentation~\cite{austern04:library_tr,sgi:_stl}.  In the
description, the variable \code{X} is used as a place holder for the
modeling type.  The \cpt{Input Iterator} concept requires several
associated types: \code{value_type}, \code{difference_type}, and
\code{iterator_category}. Associated types change from model to model.
For example, the associated \code{value_type} for \code{int*} is
\code{int} and the associated \code{value_type} for
\code{list<char>::iterator} is \code{char}. The \cpt{Input Iterator}
concept requires that the associated types be accessible via the
\code{iterator_traits} class.  (Traits classes are discussed in
Section~\ref{sec:traits}). The \code{count} algorithm, which computes
the number of occurrences of a value within a sequence, is a simple
example for the need of this access mechanism, for it needs to access
the \code{difference_type} to specify its return type:
\begin{lstlisting}
template<typename Iter, typename T>
typename iterator_traits<Iter>::difference_type
count(Iter first, Iter last, const T& value);
\end{lstlisting}
The reason that \texttt{count} uses the iterator-specific
\texttt{difference\_type} instead of \texttt{int} is to accommodate
iterators that traverse sequences that may be too long to be measured
with an \texttt{int}.

In general, a concept may consist of the following kinds of
requirements.
\begin{description}
\item[refinements] are analogous to inheritance. They allow
  one concept to include the requirements from another concept.

\item[operations] specify the functions that must be
  implemented for the modeling type.
  
\item[associated types] specify mappings between types, and in \Cpp{}
  are provided using traits classes, which we discuss in
  Section~\ref{sec:traits}.
  
\item[nested requirements] include requirements on associated types
  such as modeling a certain concept or being the same-type as another
  type.  For example, the \cpt{Input Iterator} concept requires that
  the associated \code{difference_type} be a signed integral type.

\item[semantic invariants] specify behavioral expectations about
  the modeling type.
  
\item[complexity guarantees] specify constraints on how much time or
  space may be used by an operation.
\end{description}

\begin{figure*}[p]
\begin{Sbox}
\setlength{\sidebarwidth}{\textwidth}
\addtolength{\sidebarwidth}{-2\fboxsep}
\addtolength{\sidebarwidth}{-2\fboxrule}
\addtolength{\sidebarwidth}{-2ex}
\newlength{\insidebarwidth}
\setlength{\insidebarwidth}{\sidebarwidth}
\addtolength{\insidebarwidth}{-2ex}
\begin{minipage}[t]{\sidebarwidth}

{\large \textbf{\textsf{Input Iterator}}}
\vspace{5pt}

\small
\textbf{Description}

An \cpt{Input Iterator} is an iterator that may be dereferenced to
refer to some object, and that may be incremented to obtain the next
iterator in a sequence. \cpt{Input Iterators} are not required to be
mutable. The underlying sequence elements is not required to be
persistent. For example, an \cpt{Input Iterator} could be reading
input from the terminal. Thus, an algorithm may not make multiple
passes through a sequence using an \cpt{Input Iterator}.

\vspace{5pt}
\textbf{Refinement of}

\cpt{Trivial Iterator}.

\vspace{5pt}
\textbf{Notation}

\begin{tabular}[l]{{lp{3in}}}
\texttt{X}  &     A type that is a model of \cpt{Input Iterator} \\[-1ex] 
\texttt{T}  &     The value type of \texttt{X} \\[-1ex] 
\texttt{i, j} &   Objects of type \texttt{X} \\[-1ex] 
\texttt{t}    &   Object of type \texttt{T}
\end{tabular}

\vspace{5pt}
\textbf{Associated types}
\vspace{2pt}

\begin{tabular}[l]{{|p{\insidebarwidth}|}} \hline
\code{iterator_traits<X>::value_type} \\[-1ex]
The type of the value obtained by dereferencing an \cpt{Input Iterator} \\ \hline
\code{iterator_traits<X>::difference_type} \\[-1ex]
A signed integral type used to represent the distance from one
iterator to another, or the number of elements in a range. \\ \hline
\code{iterator_traits<X>::iterator_category} \\[-1ex]
A type convertible to \code{input_iterator_tag}. \\ \hline
\end{tabular}

\vspace{5pt}
\textbf{Definitions}

An iterator is \emph{past-the-end} if it points beyond the last element of a
container. Past-the-end values are nonsingular and nondereferenceable.
An iterator is \emph{valid} if it is dereferenceable or past-the-end.
An iterator \texttt{i} is \emph{incrementable} if there is a "next" iterator, that is,
if \texttt{++i} is well-defined. Past-the-end iterators are not incrementable.
An \cpt{Input Iterator} \texttt{j} is \emph{reachable} from an \cpt{Input
  Iterator} \texttt{i} if, after applying \texttt{operator++} to \texttt{i} a finite
number of times, \texttt{i == j}.
The notation \texttt{[i,j)} refers to a range of iterators beginning
with \texttt{i} and up to but not including \texttt{j}.
The range \texttt{[i,j)} is a \emph{valid range} if both \texttt{i} and \texttt{j}
are valid iterators, and \texttt{j} is reachable from \texttt{i}.

\vspace{5pt}
\textbf{Valid expressions}

In addition to the expressions in \cpt{Trivial Iterator}, the
following expressions must be valid.

\begin{tabular}{{|l|l|l|}} \hline
\textbf{expression} & \textbf{return type} & \textbf{semantics, pre/post-conditions} \\ \hline\hline
\texttt{*i} & Convertible to \texttt{T} & pre: \texttt{i} is incrementable \\ \hline
\texttt{++i} & \code{X&} &  pre: \code{i} is dereferenceable, post: \code{i} is dereferenceable or past the end \\ \hline
\code{i++} & &  Equivalent to \code{(void)++i}. \\ \hline
\code{*i++} & & Equivalent to \code{\{T t = *i; ++i; return t;\}} \\ \hline
\end{tabular}

\vspace{5pt}
\textbf{Complexity guarantees}

All operations are amortized constant time.

\vspace{5pt}
\textbf{Models}

\code{istream_iterator}, \code{int*}, \code{list<string>::iterator}, ...

\end{minipage}
\end{Sbox}
\fbox{\TheSbox}
  \caption{Documentation for the \cpt{Input Iterator} concept.}
  \label{fig:InputIterator}
\end{figure*}

\subsection{Overview of the STL}
\label{sec:overview-stl}

\begin{figure*}[htbp]
  \centering
\includegraphics[width=4.5in]{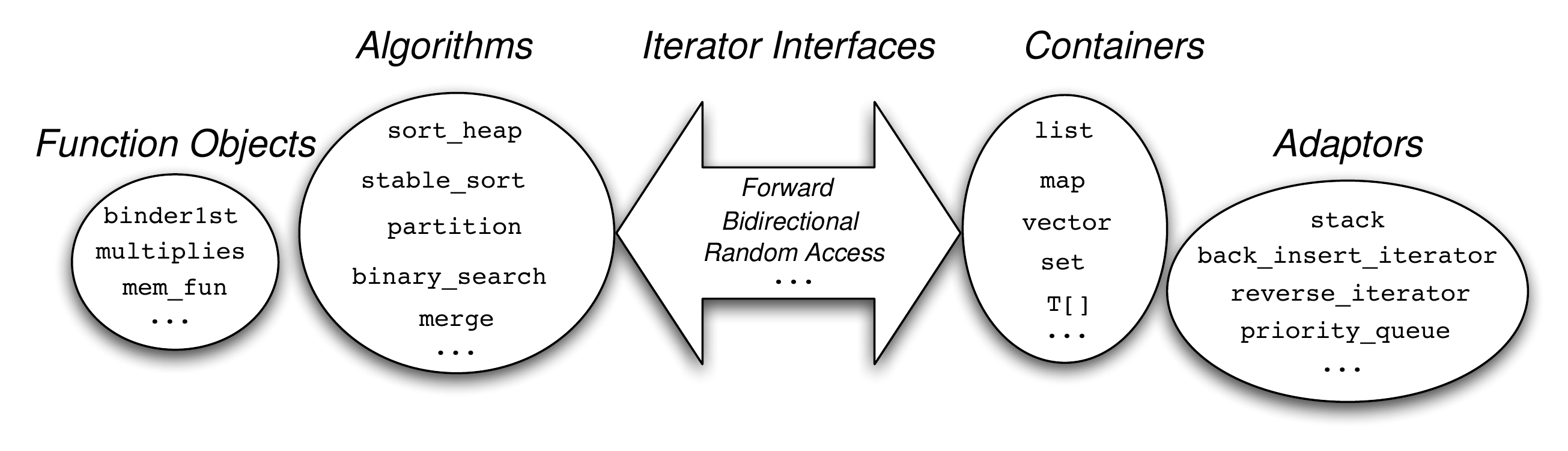}
  \caption{High-level structure of the STL.}
  \label{fig:algo-iter-data}
\end{figure*}

The high-level structure of the STL is shown in
Fig.~\ref{fig:algo-iter-data}. The STL contains over fifty generic
algorithms and 18 container classes.
The generic algorithms are implemented in terms of a family of
iterator concepts, and the containers each provide iterators that
model the appropriate iterator concepts.
As a result, the STL algorithms may be used with any of the STL
containers. In fact, the STL algorithms may be used with any data
structure that exports iterators with the required capabilities.

\begin{figure*}[hbtp]
  \centering
\includegraphics[width=5in]{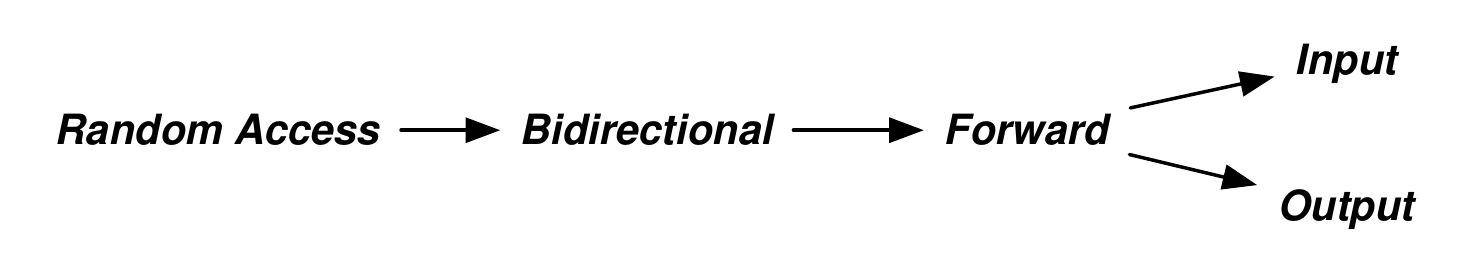}
  \caption{The refinement hierarchy of iterator concepts.}
  \label{fig:iterator-concepts}
\end{figure*}

Fig.~\ref{fig:iterator-concepts} shows the hierarchy of STL's iterator
concepts.  An arrow indicates that the source concept is a refinement
of the target. The iterator concepts arose from the requirements of
algorithms: the need to express the minimal requirements for each
algorithm.  For example, the \code{merge} algorithm passes through a
sequence once, so it only requires the basic requirements of
\concept{Input Iterator} for the two ranges it reads from and
\concept{Output Iterator} for the range it writes to.  The
\code{search} algorithm, which finds occurrences of a particular
subsequence within a larger sequence, must make multiple passes
through the sequence so it requires \cpt{Forward Iterator}.  The
\code{inplace_merge} algorithm needs to move backwards and forwards
through the sequence, so it requires \cpt{Bidirectional Iterator}.
And finally, the \code{sort} algorithm needs to jump arbitrary
distances within the sequence, so it requires \cpt{Random Access
  Iterator}. (The \code{sort} function uses the introsort
algorithm~\cite{musser97:_introsort} which is partly based on
quicksort~\cite{hoare61:_quicksort}.)
Grouping type requirements into concepts enables significant reuse of
these specifications: the \concept{Input Iterator} concept is directly
used as a type requirement in over 28 of the STL algorithms.  The
\concept{Forward Iterator}, which refines \concept{Input Iterator}, is
used in the specification of over 22 STL algorithms.

The STL includes a handful of common data structures.  When one of
these data structures does not fulfill some specialized purpose, the
programmer is encouraged to implement the appropriate specialized
data structure. All of the STL algorithms can then be made available
for the new data structure at the small cost of implementing iterators.

Many of the STL algorithms are higher-order: they take functions as
parameters, allowing the user to customize the algorithm to their own
needs. The STL defines over 25 function objects for creating and
composing functions.

The STL also contains a collection of adaptor classes, which are
parameterized classes that implement some concept in terms of the type
parameter (which is the adapted type).  For example, the
\code{back\_insert\_iterator} adaptor implements \concept{Output
  Iterator} in terms of any model of \concept{Back Insertion
  Sequence}.  The generic \code{copy} algorithm can then be used with
\code{back\_insert\_iterator} to append some
integers to a list.  Adaptors play an important role in the
plug-and-play nature of the STL and enable a high degree of reuse.

\subsection{The problem of argument dependent name lookup in C++}
\label{sec:ADL}

In \Cpp{}, uses of names inside of a template definition, such as the
use of \code{operator<} inside of \code{merge}, are resolved after
instantiation. For example, when \code{merge} is instantiated with an
iterator whose \code{value_type} is of type \code{foo::bar}, overload
resolution looks for an \code{operator<} defined for \code{foo::bar}.
If there is no such function defined in the scope of \code{merge}, 
the \Cpp{} compiler also searches the namespace where the arguments'
types are defined, so looks for \texttt{operator<} in namespace
\texttt{foo}. This rule is known as \Keyword{argument dependent
  lookup} (ADL).

The combination of implicit instantiation and ADL makes it convenient
to call generic functions. This is a nice improvement over passing
concept operations as explicit arguments to a generic function, as in
the \code{inc} example from Section~\ref{sec:introduction}.
However, ADL has two flaws. The first problem is that the programmer
calling the generic algorithm no longer has control over which
functions are used to satisfy the concept operations. Suppose that
namespace \code{foo} is a third party library and the application
programmer writing the \code{main} function has defined his own
\code{operator<} for \code{foo::bar}. ADL does not find this new
\code{operator<}.

The second and more severe problem with ADL is that it opens a hole in
the protection that namespaces are suppose to provide. ADL is applied
uniformly to all name lookup, whether or not the name is associated
with a concept in the type requirements of the template.  Thus, it is
possible for calls to helper functions to get hijacked by functions
with the same name in other namespaces. Fig.~\ref{fig:adl} shows an
example of how this can happen. The function template
\texttt{lib::generic\_fun} calls \texttt{load} with the intention of
invoking \texttt{lib::load}. In \texttt{main} we call
\texttt{generic\_fun} with an object of type \texttt{foo::bar}, so in
the call to \texttt{load}, \texttt{x} also has type \texttt{foo::bar}.
Thus, argument dependent lookup also consider namespace
\texttt{foo} when searching for \texttt{load}. There happens to be a
function named \texttt{load} in namespace \texttt{foo}, and it is a
slightly better match than \texttt{lib::foo}, so it is called instead,
thereby hijacking the call to \texttt{load}.

\begin{fdisplay}{Example problem caused by ADL.}{fig:adl}
\begin{lstlisting}
namespace lib {
  template<typename T> void load(T x, string)
    { std::cout << "Proceeding as normal!\n"; }
  template<typename T> void generic_fun(T x)
    { load(x, "file"); }
}
namespace foo {
  struct bar { int n; };
  template<typename T> void load(T x, const char*)
    { std::cout << "Hijacked!\n"; }
}
int main() {
  foo::bar a;
  lib::generic_fun(a);
}
// Output: Hijacked!
\end{lstlisting}
\end{fdisplay}

\subsection{Traits classes, template specialization, and separate type checking}
\label{sec:traits}

The traits class idiom plays an important role in writing generic
algorithms in \Cpp{}.  Unfortunately there is a deep incompatibility
between the underlying language feature, template specialization, and
our goal of separate type checking.

A traits class~\cite{myers95:_trait} maps from a type to other types
or functions.  Traits classes rely on \Cpp{} template specialization
to perform this mapping. For example, the following is the primary
template definition for
\code{iterator_traits}\index{\code{iterator_traits}}.

\begin{lstlisting}
template<typename Iterator>
struct iterator_traits { ... };
\end{lstlisting}

\noindent A specialization of \code{iterator_traits} is defined by
specifying particular type arguments for the template parameter and by
specifying an alternate body for the template.
The code below shows a user-defined iterator class, named
\code{my_iter}, and a specialization of \code{iterator_traits} for
\code{my_iter}.
\begin{lstlisting}
class my_iter {
  float operator*() { ... }
  ...
};
template<> struct iterator_traits<my_iter> {
  typedef float value_type;
  typedef int difference_type;
  typedef input_iterator_tag iterator_category;
};
\end{lstlisting}
When the type \code{iterator_traits<my_iter>} is used in other parts
of the program it refers to the above specialization. In general, a
template use refers to the most specific specialization that matches
the template arguments, if there is one, or else it refers to an
instantiation of the primary template definition.

The use of \code{iterator_traits} within a template (and template
specialization) represents a problem for separate compilation.
Consider how a compiler might type check the following
\code{unique_copy} function template.

\begin{lstlisting}
template<typename InIter, typename OutIter>
OutIter unique_copy(InIter first, InIter last, OutIter result) {
  typename iterator_traits<InIter>::value_type value = *first;
  // ...
}
\end{lstlisting}

\noindent To check the first line of the body, the compiler
needs to know that the type of \code{*first} is the same type as (or
at least convertible to) the \code{value_type} member of
\code{iterator_traits<InIter>}. However, prior to instantiation, the
compiler does not know what type \code{InIter} will be instantiated
to, and which specialization of \code{iterator_traits} to choose (and
different specializations may have different definitions of the
\code{value_type}).

Thus, if we hope to provide modular type checking, we must develop and
alternative to using traits classes for accessing associated types.

\subsection{Concept-based overloading using the tag dispatching idiom}
\label{sec:tag-dispatching}

One of the main points in the definition of generic programming in
Fig.~\ref{def:gp} is that it is sometimes necessary to provide more
than one generic algorithm for the same purpose. When this happens,
the standard approach in \Cpp{} libraries is to provide automatic
dispatching to the appropriate algorithm using the tag dispatching
idiom or \code{enable_if}~\cite{jarvi04:algorithm_specialization}.
Fig.~\ref{fig:advance} shows the \code{advance} algorithm of the STL
as it is typically implemented using the tag dispatching idiom.  The
\texttt{advance} algorithm moves an iterator forward (or backward)
\code{n} positions. There are three overloads of
\code{advance_dispatch}, each with an extra iterator tag parameter.
The \Cpp{} Standard Library defines the following iterator tag
classes, with their inheritance hierarchy mimicking the refinement
hierarchy of the corresponding concepts.

\begin{lstlisting}
struct input_iterator_tag {};
struct output_iterator_tag {};
struct forward_iterator_tag : public input_iterator_tag {};
struct bidirectional_iterator_tag : public forward_iterator_tag {};
struct random_access_iterator_tag : public bidirectional_iterator_tag {};
\end{lstlisting}

The main \code{advance} function obtains the tag for the particular
iterator from \code{iterator_traits} and then calls
\code{advance_dispatch}. Normal static overload resolution then
chooses the appropriate overload of \code{advance_dispatch}.  Both the
use of traits and the overload resolution rely on knowing actual
argument types of the template and the late type checking of \Cpp{}
templates. So the tag dispatching idiom provides another challenge for
designing a language for generic programming with separate type
checking.

\begin{fdisplay}{The \code{advance} algorithm and the tag dispatching idiom.}{fig:advance}
\begin{lstlisting}
template<typename InIter, typename Distance>
void advance_dispatch(InIter& i, Distance n, ^input_iterator_tag^) {
  while (n--) ++i;
}
template<typename BidirIter, typename Distance>
void advance_dispatch(BidirIter& i, Distance n,
                      ^bidirectional_iterator_tag^) {
  if (n > 0) while (n--) ++i;
  else while (n++) --i;
}
template<typename RandIter, typename Distance>
void advance_dispatch(RandIter& i, Distance n,
                      ^random_access_iterator_tag^) {
  i += n;
}
template<typename InIter, typename Distance>
void advance(InIter& i, Distance n) {
  ^typename iterator_traits<InIter>::iterator_category cat;^
  advance_dispatch(i, n, ^cat^);
}
\end{lstlisting}
\end{fdisplay}

\subsection{Reverse iterators and conditional models}
\label{sec:reverse-conditional-models}

The \code{reverse_iterator} class template adapts a model of
\cpt{Bidirectional Iterator} and implements \cpt{Bidirectional
  Iterator}, flipping the direction of traversal so \code{operator++}
goes backwards and \code{operator--} goes forwards. An excerpt from
the \code{reverse_iterator} class template is shown below.

\begin{lstlisting}
template<typename Iter>
class reverse_iterator {
protected:
  Iter current;
public:
  explicit reverse_iterator(Iter x) : current(x) { }
  reference operator*() const { Iter tmp = current; return *--tmp; }
  reverse_iterator& operator++() { --current; return *this; }
  reverse_iterator& operator--() { ++current; return *this; }
  reverse_iterator operator+(difference_type n) const
    { return reverse_iterator(current - n); }
  ...
}; 
\end{lstlisting}

The \code{reverse_iterator} class template is an example of a type
that models a concept conditionally: if \code{Iter} models 
\cpt{Random Access Iterator}, then so does \code{reverse_iterator<Iter>}.
The definition of
\code{reverse_iterator} defines all the operations, such as
\code{operator+}, required of a \cpt{Random Access Iterator}. The
implementations of these operations rely on the \cpt{Random Access
  Iterator} operations of the underlying \code{Iter}. One might wonder
why \code{reverse_iterator} can be used on iterators such as
\code{list<int>::iterator} that are bidirectional but not random
access. The reason this works is that a member function such as
\code{operator+} is type checked and compiled only if it is used.  For
\G{} we need a different mechanism to handle this, since function
definitions are always type checked.

\subsection{Summary of language requirements}
\label{sec:language-requirements}

In this section we surveyed how generic programming is accomplished in
\Cpp{}, taking note of the variety of language features and idioms
that are used in current practice. In this section we summarize the
findings as a list of requirements for a language to support generic
programming.

\begin{enumerate}
\item The language provides type parameterized functions with the
  ability to express constraints on the type parameters. The
  definitions of parameterized functions are type checked
  independently of how they are instantiated.
  
\item The language provides a mechanism, such as ``concepts'', for
  naming and grouping requirements on types, and a mechanism for
  composing concepts (refinement).

\item Type requirements include:
  \begin{itemize}
  \item requirements for functions and parameterized functions
  \item associated types
  \item requirements on associated types
  \item same-type constraints
  \end{itemize}
  
\item The language provides an implicit mechanism for providing
  type-specific operations to a generic function, but this mechanism
  should maintain modularity (in contrast to argument dependent lookup
  in \Cpp{}).
   
 \item The language implicitly instantiates generic
   functions when they are used.

 \item The language provides a mechanism for
   concept-based dispatching between algorithms.
   
 \item The language provides function expressions
   and function parameters.

 \item The language supports conditional modeling.
\end{enumerate}

\section{The Design of \G{}}
\label{sec:overview-g}

\G{} is a statically typed imperative language with syntax and memory
model similar to \Cpp{}. We have implemented a compiler that
translates \G{} to \Cpp{}, but \G{} could also be interpreted or compiled to
byte-code. Compilation units are separately type checked and may be
separately compiled, relying only on forward declarations from other
compilation units (even compilation units containing generic functions
and classes).
The languages features of \G{} that support generic programming are the
following:
\begin{itemize}
\item Concept and model definitions, including associated types and same-type constraints;
\item Constrained polymorphic functions, classes,
  structs, and type-safe unions;
\item Implicit instantiation of polymorphic functions; and
\item Concept-based function overloading.
\end{itemize}

\noindent In addition, \G{} includes the basic types and
control constructs \Cpp{}.

\subsection{Concepts}
\label{sec:concepts}

The following grammar defines the syntax for concepts.
\begin{lstlisting}
$\decl \leftarrow$ concept $\cid$<$\tyid,\ldots$> { $\cmem$ $\ldots$ };
$\cmem \leftarrow$ $\funsig$ $|$ $\fundef$       // Required operations
       $|$ type $\tyid$;           // Associated types
       $|$ $\type$ == $\type$;        // Same type constraints
       $|$ refines $\cid$<$\type$, $\ldots$>;
       $|$ require $\cid$<$\type$, $\ldots$>;
\end{lstlisting}
The grammar variable $\cid$ is for concept names and $\tyid$ is for
type variables.
The type variables are place holders for the modeling type (or a list of
types for multi-type concepts).
$\funsig$ and $\fundef$ are function signatures and definitions, whose
syntax we introduce later in this section.  In a concept, a function
signature says that a model must define a function with the specified
signature. A function definition in a concept provides a default
implementation.

The syntax \code{type} $\tyid$\code{;} declares an associated type; a
model of the concept must provide a type definition for the given type
name.
The syntax $\type$ \code{==} $\type$ introduces a same type
constraint.  In the context of a model definition, the two type
expressions must refer to the same type. When the concept is used in
the type requirements of a polymorphic function or class, this type
equality may be assumed. Type equality in \G{} is non-trivial, and is
explained in Section~\ref{sec:type-equality}.
Concepts may be composed with \code{refines} and \code{require}.  The
distinction is that refinement brings in the associated
types from the ``super'' concept.
Fig.~\ref{fig:input-iter-g} shows an example of a \code{concept}
definition in \G{}, the definition of \code{InputIterator}.

\begin{fdisplay}{The definition of the \concept{Input Iterator} concept in \G{}.}{fig:input-iter-g}
\begin{lstlisting}
concept InputIterator<X> {
  type value;
  type difference;
  refines EqualityComparable<X>;
  refines Regular<X>; // Regular refines Assignable and CopyConstructible
  require SignedIntegral<difference>;
  fun operator*(X b) -> value@;
  fun operator++(X! c) -> X!;
};
\end{lstlisting}
\end{fdisplay}

\subsection{Models}
\label{sec:models}

The modeling relation between a type and a concept is established with
a model definition using the following syntax.
\begin{lstlisting}
$\decl \leftarrow$ model $[$<$\tyid,\ldots$>$]$ $[$where { $\constraint$, $\ldots$ }$]$ $\cid$<$\type, \ldots$> { $\decl$ $\ldots$};  
\end{lstlisting}

\noindent The following shows an example of the \concept{Monoid}
concept and a model definition that makes \code{int} a model of
\concept{Monoid}, using addition for the binary operator and zero for
the identity element.

\begin{lstlisting}
concept Monoid<T> {
  fun identity_elt() -> T@;
  fun binary_op(T,T) -> T@;
};
model Monoid<int> {
  fun binary_op(int x, int y) -> int@ { return x + y; }
  fun identity_elt() -> int@ { return 0; }
};
\end{lstlisting}

\noindent A model definition must satisfy all requirements of the concept.
Requirements for associated types are satisfied by type definitions.
Requirements for operations may be satisfied by function definitions
in the model, by the \code{where} clause, or by functions in the
lexical scope preceding the model definition.  Refinements and nested
requirements are satisfied by preceding model definitions in the
lexical scope or by the \code{where} clause.

A model may be parameterized by placing type variables inside
\code{<>}'s after the \code{model} keyword.  The following definition
establishes that all pointer types are models of \code{InputIterator}.

\begin{lstlisting}
model <T> InputIterator<T*> {
  type value = T;
  type difference = ptrdiff_t;
};
\end{lstlisting}

\noindent The optional \code{where} clause in a model definition can be 
used to introduce constraints on the type variables. Constraints are
either modeling constraints or same-type constraints.
\begin{lstlisting}
$\constraint \leftarrow$ $\cid$<$\type$, $\ldots$> $|$ $\type$ == $\type$
\end{lstlisting}
\noindent Using the \code{where} clause we can  express
conditional modeling. As mentioned in
Section~\ref{sec:reverse-conditional-models}, we need conditional
modeling to say that \code{reverse_iterator} is a model of
\concept{Random Access Iterator} whenever the underlying iterator is.
Fig.~\ref{fig:reverse-models-rand-access} shows is a model definition
that says just this.

\begin{fdisplay}{\code{reverse_iterator} conditionally models
  the \concept{Random Access Iterator} concept.}{fig:reverse-models-rand-access}
\begin{lstlisting}
model <Iter> where { RandomAccessIterator<Iter> }
RandomAccessIterator< reverse_iterator<Iter> >
{
  fun operator+(reverse_iterator<Iter> r, difference n)
     -> reverse_iterator<Iter>@
    { return @reverse_iterator<Iter>(r.current + n); }
  fun operator-(reverse_iterator<Iter> r, difference n) 
    -> reverse_iterator<Iter>@
    { return @reverse_iterator<Iter>(r.current - n); }
  fun operator-(reverse_iterator<Iter> a, reverse_iterator<Iter> b)
     -> difference
    { return a.current - b.current; }
};
\end{lstlisting}
\end{fdisplay}

The rules for type checking parameterized model definitions with
constraints is essentially the same as for generic functions, which we
discuss in Section~\ref{sec:generic-functions}.

\subsection{Nominal versus structural conformance}

One of the fundamental design choices of \G{} was to include model
definitions. After all, it is possible to instead have the compiler
figure out when a type has implemented all of the requirements of a
concept. We refer to the approach of using explicit model definitions
\Keyword{nominal conformance} whereas the implicit approach we call
\Keyword{structural conformance}. An example of the nominal versus
structural distinction can be seen in the example below.  Do the
concepts create two ways to refer to the same concept or are they
different concepts that happen to have the same constraints?

\begin{tabular}[l]{p{2.5in}p{2.5in}}
\begin{lstlisting}
concept A<T> {
  fun foo(T x) -> T;
};
\end{lstlisting}
&
\begin{lstlisting}
concept B<T> {
  fun foo(T x) -> T;
};
\end{lstlisting}
\end{tabular}

\noindent With nominal conformance, the above are two different
concepts, whereas with structural conformance, \code{A} and \code{B}
are two names for the same concept.  Examples of language mechanisms
providing nominal conformance include Java interfaces\index{interface}
and Haskell type classes\index{type class}. Examples of language
mechanisms providing structural conformance include ML
signatures\index{signature}~\cite{milner90:definition_sml}, Objective
Caml\index{Objective Caml} object types\index{object
  types}~\cite{leroy03:_ocaml}, CLU\index{CLU} type sets\index{type
  sets}~\cite{liskov79:_clu_ref}, and Cforall
specifications\index{Cforall}~\cite{ditchfield96:_overview_cforall}.

Choosing between nominal and structural conformance is difficult
because both options have good arguments in their favor.

\textbf{Structural conformance is more convenient than nominal conformance} 
With nominal conformance, the modeling relationship is established by
an explicit declaration. For example, a Java class declares that it
\code{implements} an interface. In Haskell, an \code{instance}
declaration establishes the conformance between a particular type and
a type class. When the compiler sees the explicit declaration, it
checks whether the modeling type satisfies the requirements of the
concept and, if so, adds the type and concept to the modeling
relation.

Structural conformance, on the other hand, requires no explicit
declarations. Instead, the compiler determines on a need-to-know
basis whether a type models a concept. The advantage is that
programmers need not spend time writing explicit declarations.

\textbf{Nominal conformance is safer than structural conformance}
The usual argument against structural conformance is that it is prone
to \Keyword{accidental conformance}. The classic example of this is a
cowboy object being passed to something expecting a
\code{Window}~\cite{magnusson91:cowboy_rectangle}. The \code{Window}
interface includes a \code{draw()} method, which the cowboy has, so
the type system does not complain even though something wrong has
happened.  This is not a particularly strong argument because the
programmer has to make a big mistake for this kind accidental
conformance to occur.

However, the situation changes for languages that support
concept-based overloading. For example, in
Section~\ref{sec:tag-dispatching} we discussed the tag-dispatching
idiom used in \Cpp{} to select the best \code{advance} algorithm
depending on whether the iterator type models \cpt{Random Access
  Iterator} or only \cpt{Input Iterator}.
With concept-based overloading, it becomes possible for accidental
conformance to occur without the programmer making a mistake. The
following \Cpp{} code is an example where an error would occur if
structural conformance were used instead of nominal.

\begin{lstlisting}
std::vector<int> v;
std::istream_iterator<int> in(std::cin), in_end;
v.insert(v.begin(), in, in_end);
\end{lstlisting}

The \code{vector} class has two versions of \code{insert}, one for
models of \cpt{Input Iterator} and one for models of \cpt{Forward
  Iterator}. An \cpt{Input Iterator} may be used to traverse a range
only a single time, whereas a \cpt{Forward Iterator} may traverse
through its range multiple times. Thus, the version of \code{insert}
for \cpt{Input Iterator} must resize the vector multiple
times as it progresses through the input range. In contrast, the
version of \code{insert} for \cpt{Forward Iterator} is more efficient
because it first discovers the length of the range (by calling
\code{std::distance}, which traverses the input range), resizes the
vector to the correct length, and then initializes the vector from the
range.

The problem with the above code is that \code{istream_iterator}
fulfills the syntactic requirements for a \cpt{Forward Iterator} but
not the semantic requirements: it does not support multiple passes.
That is, with structural conformance, there is a false positive and
\code{insert} dispatches to the version for \cpt{Forward Iterator}s.
The program resizes the vector to the appropriate size for all the
input but it does not initialize the vector because all of the input
has already been read.

\textbf{Why not both?}
It is conceivable to provide both nominal and structural conformance
on a concept-by-concept basis, which is in fact the approach used in
the concept extension for \Cpp{}0X. Concepts that are intended to be
used for dispatching could be nominal and other concepts could be
structural. This matches the current \Cpp{} practice: some concepts
come with traits classes that provide nominal conformance whereas
other concepts do not (the default situation with \Cpp{} templates is
structural conformance). However, providing both nominal conformance
and structural conformance complicates the language, especially for
programmers new to the language, and degrades its uniformity.
Therefore, with \G{} we provide only nominal conformance, giving
priority to safety and simplicity over convenience.

\subsection{Generic Functions}
\label{sec:generic-functions}

The syntax for generic functions is shown below.  The name of the
function is the identifier after \code{fun}, the type parameters are
between the \code{<>}'s and are constrained by the requirement in the
\code{where} clause. A function's parameters are between the
\code{()}'s and the return type of a function comes after the
\code{->}.
\begin{lstlisting}
$\fundef \leftarrow$ fun $\id$ $[$<$\tyid,\ldots$>$]$ $[$where { $\constraint$, $\ldots$ }$]$
                ($\type$ $\pass$ $[\id]$, $\ldots$) -> $\type$ $\pass$ { $\stmt$ $\ldots$ }
$\funsig \leftarrow$ fun $\id$ $[$<$\tyid,\ldots$>$]$ $[$where { $\constraint$, $\ldots$ }$]$
              ($\type$ $\pass$ $[\id]$, $\ldots$) -> $\type$ $\pass$;
$\decl \leftarrow$ $\fundef$ $|$ $\funsig$
$\pass \leftarrow$ $\mathit{mut}$ $\mathit{ref}$     // pass by reference
      $|$ @           // pass by value
$\mathit{mut} \leftarrow$ const $|$ $\epsilon$    // constant
      $|$ !           // mutable 
$\mathit{ref} \leftarrow$ & $|$ $\epsilon$
\end{lstlisting}

\noindent The default parameter passing mode in \G{} is read-only
pass-by-reference. Read-write pass-by-reference is indicated by
\code{!} and pass-by-value is indicated by \code{@}.

The \code{merge} algorithm, implemented as a generic function in \G{},
is shown in Fig.~\ref{fig:merge-g}. The function is parameterized on
three types: \code{Iter1}, \code{Iter2}, and \code{Iter3}.  The dot
notation is used to refer to a member of a model, including associated
types such as the \code{value} type of an iterator.
\begin{lstlisting}
$\assoc \leftarrow$ $\cid$<$\type$, $\ldots$>.$\id$ $|$ $\cid$<$\type$, $\ldots$>.$\assoc$
$\type \leftarrow$ $\assoc$
\end{lstlisting}
The \concept{Output Iterator} concept used in the \code{merge} function
is an example of a multi-parameter concept. It has a type parameter
\code{X} for the iterator and a type parameter \code{T} for the type
that can be written to the iterator. The following is the definition
of the \concept{Output Iterator} concept.

\begin{lstlisting}
concept OutputIterator<X,T> {
  refines Regular<X>;
  fun operator<<(X! c, T t) -> X!;
};
\end{lstlisting}

\begin{fdisplay}{The \code{merge} algorithm in \G{}.}{fig:merge-g}
\begin{lstlisting}
fun merge<Iter1,Iter2,Iter3>
where { InputIterator<Iter1>, InputIterator<Iter2>,
        LessThanComparable<InputIterator<Iter1>.value>,
        InputIterator<Iter1>.value == InputIterator<Iter2>.value,
        OutputIterator<Iter3, InputIterator<Iter1>.value> }
(Iter1@ first1, Iter1 last1, Iter2@ first2, Iter2 last2, Iter3@ result)
  -> Iter3@
{
  while (first1 != last1 and first2 != last2) {
    if (*first2 < *first1) {
      result << *first2; ++first2;
    } else {
      result << *first1; ++first1;
    }
  }
  return copy(first2, last2, copy(first1, last1, result));
}
\end{lstlisting}
\end{fdisplay}

In general the body of a generic function contains a sequence of
statements. Syntax for some of the statements in \G{} is defined in
the following grammar.
\begin{lstlisting}
$\stmt \leftarrow$ let $\id$ = $\expr$; $|$ while ($\expr$) $\stmt$ $|$ return $\expr$; $|$ $\expr$; $|$ $\ldots$
\end{lstlisting}
The \code{let} form introduces local variables, deducing the type of
the variable from the right-hand side expression (similar to the
\code{auto} proposal for
\Cpp{}0X~\cite{jarvi:03:std_decltype_proposal}).

The body of a generic function is type checked separately from any
instantiation of the function. The type parameters are treated as
abstract types so no type-specific operations may be applied to them
unless otherwise specified by the \code{where} clause. The
\code{where} clause introduces surrogate model definitions and
function signatures (for all the required concept operations) into the
scope of the function. 

Multiple functions with the same name may be defined, and static
overload resolution is performed by \G{} to decide which function to
invoke at a particular call site depending on the argument types and
also depending on which model definitions are in scope.  When more
than one overload may be called, the most specific overload is called
(if one exists) according to the rules described in
Section~\ref{sec:concept-based-overloading}.

\subsection{Function calls and implicit instantiation}
\label{sec:implicit-instantiation}

The syntax for calling functions (or polymorphic functions) is the
C-style notation:

\begin{lstlisting}
$\expr \leftarrow$ $\expr$($\expr$, $\ldots$)
\end{lstlisting}

\noindent Arguments for the type parameters of a polymorphic
function need not be supplied at the call site: \G{} will deduce the
type arguments by unifying the types of the arguments with the types
of the parameters and then implicitly instantiate the polymorphic
function. The design issues surrounding implicit instantiation are
described below. All of the requirements in the \code{where} clause
must be satisfied by model definitions in the lexical scope preceding
the function call, as described in Section~\ref{sec:model-lookup}. The
following is an example of calling the generic \code{accumulate}
function.  In this case, the generic function is implicitly
instantiated with type argument \code{int*}.

\begin{lstlisting}
fun main() -> int@@ {
  let a = new int[8];
  a[0] = 1; a[1] = 2; a[2] = 3; a[3] = 4; a[4] = 5; 
  let s = accumulate(a, a + 5);
  if (s == 15) return 0;
  else return -1;
}  
\end{lstlisting}

\noindent A polymorphic function may be explicitly instantiated
using this syntax:

\begin{lstlisting}
$expr \leftarrow$ $expr$<|$ty$, $\ldots$|>
\end{lstlisting}

Following Mitchell~\cite{mitchell88:_poly_inf_containment} we view
implicit instantiation as a kind of coercion that transforms an
expression of one type to another type. In the example above, the
\code{accumulate} function was coerced from
\begin{lstlisting}
fun <Iter> where
  { InputIterator<Iter>, Monoid<InputIterator<Iter>.value> }
  (Iter@, Iter) -> InputIterator<Iter>.value@
\end{lstlisting}
to
\begin{lstlisting}
fun (int*@, int*) -> InputIterator<int*>.value@
\end{lstlisting}
There are several kinds of implicit coercions in \G{}, and together
they form a subtyping relation $\leq$.  The subtyping relation is
reflexive and transitive. Like \Cpp{}, \G{} contains some
bidirectional implicit coercions, such as \code{float} $\leq$
\code{double} and \code{double} $\leq$ \code{float}, so $\leq$ is not
anti-symmetric.
The subtyping relation for \G{} is defined by a set of subtyping
rules.
The following is the subtyping rule for generic function
instantiation.
\begin{equation*}
\inference[(\textsc{Inst})]
  {\Gamma \text{ satisfies } \rep{c}}
  {\Gamma |- \fun\LT\rep{\alpha}\GT \where \LB \rep{c} \RB \LP\rep{\sigma}\RP\arrow \tau
    \leq [\rep{\rho}/\rep{\alpha}] (\fun\LP\rep{\sigma}\RP\arrow\tau)}
\end{equation*}
The type parameters $\rep{\alpha}$ are substituted for type arguments
$\rep{\rho}$ and the constraints in the \code{where} clause must be
satisfied in the current environment. To apply this rule, the compiler
must choose the type arguments. We call this \Keyword{type argument
  deduction} and discuss it in more detail momentarily.  Constraint
satisfaction is discussed in Section~\ref{sec:model-lookup}.

The subtyping relation allows for coercions during type checking
according to the subsumption rule:
\begin{equation*}
 \inference[(\textsc{Sub})]
 {\Gamma |- e : \sigma
  & \Gamma |- \sigma \leq \tau}
{\Gamma |- e : \tau}
\end{equation*}
The (\textsc{Sub}) rule is not syntax-directed so its addition to the
type system would result in a non-deterministic type checking
algorithm. The standard workaround is to omit the above rule and
instead allow coercions in other rules of the type system such as the
rule for function application. The following is a rule for function
application that allows coercions in both the function type and in the
argument types.
\begin{equation*}
  \inference[(\textsc{App})]
   {\Gamma |- e_1 : \tau_1
    & \Gamma |- \rep{e_2} : \rep{\sigma_2}
    & \Gamma |- \tau_1 \leq \fun\LP\rep{\sigma_3}\RP\arrow \tau_2
    & \Gamma |- \rep{\sigma_2} \leq \rep{\sigma_3}
   }
   {\Gamma |- e_1(\rep{e_2}) : \tau_2}
\end{equation*}

As mentioned above, the type checker must guess the type arguments
$\rep{\rho}$ to apply the (\textsc{Inst}) rule. In addition, the
(\textsc{App}) rule includes several types that appear from nowhere:
$\rep{\sigma_3}$ and $\rep{\tau_2}$. The problem of deducing these
types is equivalent to trying to find solutions to a system of
inequalities. Consider the following example program.
\begin{lstlisting}
fun apply<T>(fun(T)->T f, T x) -> T { return f(x); }
fun id<U>(U a) -> U { return a; }
fun main() -> int@ { return apply(id, 0); }
\end{lstlisting}
The application \code{apply(id, 0)} type checks if there is a solution
to the following system:
\begin{lstlisting}
fun<T>(fun(T)->T, T) -> T §$\leq$§ fun(§$\alpha$§, §$\beta$§) -> §$\gamma$§
fun<U>(U)->U §$\leq$§ §$\alpha$§
int §$\leq$§ §$\beta$§
\end{lstlisting}
The following type assignment is a solution to the above system.
\begin{lstlisting}
§$\alpha$§ §$=$§ fun(int)->int
§$\beta$§ §$=$§ int
§$\gamma$§ §$=$§ int
\end{lstlisting}

Unfortunately, not all systems of inequalities are as easy to solve as
the above example.  In fact, with Mitchell's original set of subtyping
rules, the problem of solving systems of inequalities was proved
undecidable by Tiuryn and
Urzyczyn~\cite{tiuryn02:subtyping_2nd_order}. There are several
approaches to dealing with this undecidability.

\textbf{Remove the (\textsc{Arrow}) rule.}
Mitchell's subtyping relation included the usual co/contravariant rule
for functions.
\begin{equation*}
\inference[(\textsc{Arrow})]
  {\rep{\sigma_2} \leq \rep{\sigma_1} & \tau_1 \leq \tau_2}
  {\fun\LP\rep{\sigma_1}\RP\arrow \tau_1 \leq \fun\LP\rep{\sigma_2}\RP\arrow \tau_2}
\end{equation*}
The (\textsc{Arrow}) rule is nice to have because it allows a function
to be coerced to a different type so long as the parameter and return
types are coercible in the appropriate way. In the following example
the standard \code{ilogb} function is passed to \code{foo} even though
it does not match the expected type. The (\textsc{Arrow}) rule allows
for this coercion because \code{int} is coercible to \code{double}.
\begin{lstlisting}
include "math.h"; // fun ilogb(double x) -> int;
fun foo(fun(int)->int@ f) -> int@ { return f(1); }
fun main() -> int@ { return foo(ilogb); }
\end{lstlisting}

However, the (\textsc{Arrow}) rule is one of the culprits in the
undecidability of the subtyping problem; removing it makes the problem
decidable~\cite{tiuryn02:subtyping_2nd_order}.
The language \MLF of Le Botlan and Remy~\cite{lebotlan-remy!mlf-icfp}
takes this approach, and for the time being, so does \G{}.  With this
restriction, type argument deduction is reduced to the variation of
unification defined in~\cite{lebotlan-remy!mlf-icfp}.  Instead of
working on a set of variable assignments, this unification algorithm
keeps track of either a type assignment or the tightest lower bound
seen so far for each variable.  The (\textsc{App}) rule for \G{} is
reformulated as follows to use this \code{unify} algorithm.

\begin{equation*}
  \inference[(\textsc{App})]
   {\Gamma |- e_1 : \tau_1
    & \Gamma |- \rep{e_2} : \rep{\sigma_2} \\
     Q =  \{ \tau_1 \leq \alpha, \rep{\sigma_2} \leq \rep{\beta} \} 
    & Q' = \mathtt{unify}(\alpha, \fun\LP\rep{\beta}\RP \arrow \gamma, Q)
    }
   {\Gamma |- e_1(\rep{e_2}) : Q'(\gamma)}
\end{equation*}

In languages where functions are often written in curried form, it is
important to provide even more flexibility than in the above
(\textsc{App}) rule by postponing instantiation, as is done in \MLF.
Consider the \code{apply} example again, but this time written in
curried form.
\begin{lstlisting}
fun apply<T>(fun(T)->T f) -> (fun(T)->T)@ {
  return fun(T x) { return f(x); };
}
fun id<U>(U a) -> U { return a; }
fun main() -> int@ { return apply(id)(0); }
\end{lstlisting}
In the first application \code{apply(id)} we do not yet know that
\code{T} should be bound to \code{int}. The instantiation needs to be
delayed until the second application \code{apply(id)(0)}.  In general,
each application contributes to the system of inequalities that needs
to be solved to instantiate the generic function.  In \MLF, the return
type of each application encodes a partial system of inequalities.
The inequalities are recorded in the types as lower bounds on type
parameters. The following is an example of such a type.
\begin{lstlisting}
fun<U> where { fun<T>(T)->T $\leq$ U } (U) -> U
\end{lstlisting}
Postponing instantiation is not as important in \G{} because functions
take multiple parameters and currying is seldom used.

Removal of the arrow rule means that, in some circumstances, the
programmer would have to wrap a function inside another function before
passing the function as an argument.

\textbf{Restrict the language to predicative polymorphism}
Another alternative is to restrict the language so that only monotypes
(non-generic types) may be used as the type arguments in an
instantiation. This approach is used in by Odersky and
L\"aufer~\cite{odersky96:_putting} and also by Peyton Jones and
Shields~\cite{jones04:_type_inf_arb_rank}.  However, this approach
reduces the expressiveness of the language for the sake of the
convenience of implicit instantiation.

\textbf{Restrict the language to second-class polymorphism}
Restricting the language of types to disallow polymorphic types nested
inside other types is another way to make the subtyping problem
decidable. With this restriction the subtyping problem is solved
by normal unification. Languages such as SML and Haskell 98 use this
approach.  Like the restriction to predicative polymorphism, this
approach reduces the expressiveness of the language for the sake of
implicit instantiation (and type inference).  However, there are many
motivating use cases for first-class
polymorphism~\cite{shan04:_sexy_types}, so throwing out first-class
polymorphism is not our preferred alternative.

\textbf{Use a semi-decision procedure}
Yet another alternative is to use a semi-decision procedure for the
subtyping problem. The advantage of this approach is that it allows
implicit instantiation to work in more situations, though it is not
clear whether this extra flexibility is needed in practice. The down
side is that there are instances of the subtyping problem where the
procedure diverges and never returns with a solution.

\subsection{Model lookup (constraint satisfaction)}
\label{sec:model-lookup}

The basic idea behind model lookup is simple although some of the
details are a bit complicated.  Consider the following program
containing a generic function \code{foo} with a requirement for
\code{C<T>}.
\begin{lstlisting}
concept C<T> { };
model C<int> { };
\end{lstlisting}
\begin{lstlisting}
fun foo<T> where { C<T> } (T x) -> T { return x; }

fun main() -> int@ {
  return foo(0);// lookup model C<int>
}
\end{lstlisting}
At the call \code{foo(0)}, the compiler deduces the binding
\code{T=int} and then seeks to satisfy the \code{where} clause, with
\code{int} substituted for \code{T}. In this case the constraint
\code{C<int>} must be satisfied. In the scope of the call
\code{foo(0)} there is a model definition for \code{C<int>}, so the
constraint is satisfied. We call \code{C<int>} the \Keyword{model
  head}.

\subsubsection{Lexical scoping of models}
\label{sec:scoped-models}

The design choice to look for models in the lexical scope of the
instantiation is an important choice for \G{}, and differentiates it
from both Haskell and the concept extension for \Cpp{}. This choice
improves the modularity of \G{} by preventing model declarations in
separate modules from accidentally conflicting with one another.

For example, in Fig.~\ref{fig:overlapping-models} we create \code{sum}
and \code{product} functions in modules \code{A} and \code{B}
respectively by instantiating \code{accumulate} in the presence of
different model declarations.  This example would not type check in
Haskell, even if the two instance declarations were to be placed in
different modules, because instance declarations implicitly leak out
of a module when anything in the module is used by another module.
This example would be illegal in \Cpp{}0X concept extension because 1)
model definitions must appear in the same namespace as their concept,
and 2) if placed in the same namespace, the two model definitions
would violate the one-definition-rule.

\begin{fdisplay}{Intentionally overlapping models.}{fig:overlapping-models}
\begin{lstlisting}
module A {
  model Monoid<int> {
    fun binary_op(int x, int y) -> int@ { return x + y; }
    fun identity_elt() -> int@ { return 0; }
  };
  fun sum<Iter>(Iter first, Iter last) -> int {
    return accumulate(first, last);
  }
}

module B {
  model Monoid<int> {
    fun binary_op(int x, int y) -> int@ { return x * y; }
    fun identity_elt() -> int@ { return 1; }
  };
  fun product<Iter>(Iter first, Iter last) -> int {
    return accumulate(first, last);
  }
}
\end{lstlisting}
\end{fdisplay}

It is also quite possible for separately developed modules to include
model definitions that accidentally overlap. In \G{}, this is not a
problem, as the model definitions will each apply within their own
module. Model definitions may be explicitly imported from one module
to another. The syntax for modules and import declarations is shown
below. An interesting extension would be parameterized modules, but we
leave that for future work.

\begin{lstlisting}
$\decl$ $\leftarrow$ module $\moduleid\;\LB\;\decl \ldots \; \RB$ // module
  $|$ scope $\moduleid$ = $\mathit{scope}$; // scope alias 
  $|$ import $\mathit{scope}$.$\cid$<$\type$,$\ldots$>; // import model 
  $|$ public: $\decl \ldots$ // public region 
  $|$ private: $\decl \ldots$ // private region 
\end{lstlisting}

\subsubsection{Constrained models}

In \G{}, a model definition may itself be parameterized and the type
parameters constrained by a \code{where} clause.
Fig.~\ref{fig:param-model} shows a typical example of a parameterized
model. The model definition in the example says that for any type
\code{T}, \code{list<T>} is a model of \code{Comparable} if \code{T}
is a model of \code{Comparable}. Thus, a model definition is like an
inference rule or a Horn clause~\cite{horn51:_sentences} in logic
programming. For example, a model definition of the form
\begin{lstlisting}
model <T1,...,Tn> where { P1, ..., Pn }
Q { ... };
\end{lstlisting}
corresponds to the Horn clause:
\begin{equation*}
(P_1 \text{ and } \ldots \text{ and } P_n) \text{ implies } Q
\end{equation*}
The model definitions from the example in Fig.~\ref{fig:param-model} could
be represented in Prolog with the following two rules:
\begin{lstlisting}
comparable(int).
comparable(list(T)) :- comparable(T).
\end{lstlisting}

\begin{fdisplay}{Example of parameterized model definition.}{fig:param-model}
  \begin{lstlisting}
concept Comparable<T> {
  fun operator==(T,T)->bool@;
};
model Comparable<int> { };

struct list<T> { /*...*/ };

model <T> where { Comparable<T> }
Comparable< list<T> > {
  fun operator==(list<T> x, list<T> y) -> bool@ { /*...*/ }
};

fun generic_foo<C> where { Comparable<C> } (C a, C b) -> bool@
  { return a == b; }

fun main() -> int@ {
  let l1 = @list<int>(); let l2 = @list<int>();
  generic_foo(l1,l2);
  return 0;
}
  \end{lstlisting}
\end{fdisplay}

The algorithm for model lookup is essentially a logic programming
engine: it performs unification and backward chaining (similar to how
instance lookup is performed in Haskell).
Unification is used to determine when the head of a model definition
matches. For example, in Fig.~\ref{fig:param-model}, in the call to
\code{generic\_foo} the constraint \code{Comparable< list<int> >}
needs to be satisfied. There is a model definition for
\code{Comparable< list<T> >} and unification of \code{list<int>} and
\code{list<T>} succeeds with the type assignment \code{T} $=$
\code{int}.
However, we have not yet satisfied \code{Comparable< list<int> >}
because the \code{where} clause of the parameterized model must also
be satisfied. The model lookup algorithm therefore proceeds
recursively and tries to satisfy \code{Comparable<int>}, which in this
case is trivial.  This process is called \Keyword{backward chaining}:
it starts with a goal (a constraint to be satisfied) and then applies
matching rules (model definitions) to reduce the goal into subgoals.
Eventually the subgoals are reduced to facts (model definitions
without a \code{where} clause) and the process is complete. As is
typical of Prolog implementations, \G{} processes subgoals in a
depth-first manner.

It is possible for multiple model definitions to match a constraint.
When this happens the most specific model definition is used, if one
exists. Otherwise the program is ill-formed. We say that definition
$A$ is a \Keyword{more specific model} than definition $B$ if the head
of $A$ is a substitution instance of the head of $B$ and if the
\code{where} clause of $B$ implies the \code{where} clause of $A$.  In
this context, implication means that for every constraint $c$ in the
\code{where} clause of $A$, $c$ is satisfied in the current
environment augmented with the assumptions from the \code{where}
clause of $B$.

\G{} places very few restrictions on the form of a model definition.
The only restriction is that all type parameters of a model must
appear in the head of the model. That is, they must appear in the type
arguments to the concept being modeled. For example, the following
model definition is ill formed because of this restriction.
\begin{lstlisting}
concept C<T> { };
model <T,U> C<T> { }; // ill formed, U is not in an argument to C
\end{lstlisting}
This restriction ensures that unifying a constraint with the model
head always produces assignments for all the type parameters.

Horn clause logic is by nature powerful enough to be Turning-complete.
For example, it is possible to express general recursive functions.
The program in Fig.~\ref{fig:ack} computes the Ackermann function at
compile time by encoding it in model definitions. This power comes at
a price: determining whether a constraint is satisfied by a set of
model definitions is in general undecidable. Thus, model lookup is not
guaranteed to terminate and programmers must take some care in writing
model definitions.  We could restrict the form of model definitions to
achieve decidability however there are two reasons not to do so.
First, restrictions would complicate the specification of \G{} and
make it harder to learn.  Second, there is the danger of ruling out
useful model definitions.

\begin{fdisplay}{The Ackermann function encoded in model definitions.}{fig:ack}
  \begin{lstlisting}
struct zero { };
struct suc<n> { };
concept Ack<x,y> { type result; };

model <y> Ack<zero,y> { type result = suc<y>; };

model <x> where { Ack<x, suc<zero> > }
Ack<suc<x>, zero> { type result = Ack<x, suc<zero> >.result; };

model <x,y> where { Ack<suc<x>,y>, Ack<x, Ack<suc<x>,y>.result > }
Ack< suc<x>,suc<y> > {
  type result = Ack<x, Ack<suc<x>,y>.result >.result;
};

fun foo(int) { }
fun main() -> int@ {
  type two = suc< suc<zero> >; type three = suc<two>;
  foo(@Ack<two,three>.result());
  // error: Type (suc<suc<suc<suc<suc<suc<suc<suc<suc<zero>>>>>>>>>) 
  // does not match type (int)
}
  \end{lstlisting}
\end{fdisplay}

\subsection{Improved error messages}

In the introduction we showed how users of generic libraries in \Cpp{}
are plagued by hard to understand error messages.  The introduction of
concepts and where clauses in \G{} solves this problem. The following
is the same misuse of the \code{stable\_sort} function, but this time
written in \G{}.
\begin{lstlisting}[numbers=left,stepnumber=1,firstnumber=4]
fun main() -> int@{
  let v = @list<int>();
  stable_sort(begin(v), end(v));
  return 0;
}
\end{lstlisting}
In contrast to long \Cpp{} error message
(Fig.~\ref{fig:stable-sort-error}), in \G{} we get the following:

\begin{lstlisting}[deletekeywords={fun,where,concept,model,require,refines,type,let,module,scope,import,not,int},]
test/stable_sort_error.hic:6:
In application stable_sort(begin(v), end(v)),
Model MutableRandomAccessIterator<mutable_list_iter<int>> 
needed to satisfy requirement, but it is not defined.
\end{lstlisting}

A related problem that plagues authors of generic \Cpp{} libraries is
that type errors often go unnoticed during library development. Again,
this is because \Cpp{} delays type checking templates until
instantiation. One of the reasons for such type errors is that the
implementation of a template is not consistent with its documented
type requirements.

This problem is directly addressed in \G{}: the implementation of a
generic function is type-checked with respect to its \code{where}
clause, independently of any instantiations. Thus, when a generic
function successfully compiles, it is guaranteed to be free of type
errors and the implementation is guaranteed to be consistent with the
type requirements in the \code{where} clause.

Interestingly, while implementing the STL in \G{}, the type checker
caught several errors in the STL as defined in \Cpp{}.
One such error was in \code{replace\_copy}. The implementation below
was translated directly from the GNU \Cpp{} Standard Library, with the
\code{where} clause matching the requirements
for \code{replace\_copy} in the \Cpp{} Standard~\cite{cpp98}.

\begin{lstlisting}[numbers=left,stepnumber=1,firstnumber=196]
fun replace_copy<Iter1,Iter2, T>
where { InputIterator<Iter1>, Regular<T>, EqualityComparable<T>,
        OutputIterator<Iter2, InputIterator<Iter1>.value>,
        OutputIterator<Iter2, T>,
        EqualityComparable2<InputIterator<Iter1>.value,T> }
(Iter1@ first, Iter1 last, Iter2@ result, T old, T neu) -> Iter2@ {
  for ( ; first != last; ++first)
    result << *first == old ? neu : *first;
  return result;
}
\end{lstlisting}

\noindent The \G{} compiler gives the following error message:
\begin{lstlisting}
stl/sequence_mutation.hic:203:
The two branches of the conditional expression must have the 
same type or one must be coercible to the other.
\end{lstlisting}

\noindent This is a subtle bug, which explains why it has
gone unnoticed for so long. The type requirements say that both the
value type of the iterator and \code{T} must be writable to the
output iterator, but the requirements do not say that the value type
and \code{T} are the same type, or coercible to one another.

\subsection{Generic classes, structs, and unions}

The syntax for generic classes, structs, and unions is defined below.
The grammar variable $\clid$ is for class, struct, and union names.

\begin{lstlisting}
$\decl \leftarrow$ class $\clid$ $\polyhdr$ { $\classmem$ $\ldots$ };
$\decl \leftarrow$ struct $\clid$ $\polyhdr$ { $\mem$ $\ldots$ };
$\decl \leftarrow$ union $\clid$ $\polyhdr$ { $\mem$ $\ldots$ };
$\mem \leftarrow$ $\type$ $\id$;
$\classmem \leftarrow$ $\mem$
         $|$ $\polyhdr$ $\clid$($\type$ $\pass$ $[\id]$, $\ldots$) { $\stmt$ $\ldots$ }
         $|$ ~$\clid$() { $\stmt$ $\ldots$ }
$polyhdr \leftarrow$ $[$<$\tyid,\ldots$>$]$ $[$where { $\constraint$, $\ldots$ }$]$
\end{lstlisting}

\noindent Classes consist of data members, constructors, and a destructor.
There are no member functions; normal functions are used instead.
Data encapsulation (\code{public}/\code{private}) is specified at the
module level instead of inside the class. Class, struct, and unions
are used as types using the syntax below. Such a type is well-formed
if the type arguments are well-formed and if the requirements in its
where clause are satisfied.

\begin{lstlisting}
$\type \leftarrow$ $\clid$$[$<$\type$, $\ldots$>$]$
\end{lstlisting}

\subsection{Type equality}
\label{sec:type-equality}

There are several language constructions in \G{} that make it
difficult to decide when two types are equal. Generic functions
complicate type equality because the names of the type parameters do
not matter. So, for example, the following two function types are
equal:
\begin{lstlisting}
fun<T>(T)->T $=$ fun<U>(U)->U
\end{lstlisting}
The order of the type parameters does matter (because a generic
function may be explicitly instantiated) so the following two types
are not equal.
\begin{lstlisting}
fun<S,T>(S,T)->T $\neq$ fun<T,S>(S,T)->T
\end{lstlisting}
Inside the scope of a generic function, type parameters with different
names are assumed to be different types (this is a conservative
assumption). So, for example, the following program is ill formed
because variable \code{a} has type \code{S} whereas function \code{f} is
expecting an argument of type \code{T}.
\begin{lstlisting}
fun foo<S, T>(S a, fun(T)->T f) -> T { return f(a); }
\end{lstlisting}

Associated types and same-type
constraints also affect type equality.
First, if there is a model definition in the current scope such as:
\begin{lstlisting}
model C<int> { type bar = bool; };
\end{lstlisting}
then we have the equality \code{C<int>.bar} $=$ \texttt{bool}.

Inside the scope of a generic function, same-type constraints help
determine when two types are equal. For example, the following version of
\code{foo} is well formed:
\begin{lstlisting}
fun foo_1<T, S> where { T == S } (fun(T)->T f, S a) -> T { return f(a); }
\end{lstlisting}
\noindent There is a subtle difference between the above version of
\code{foo} and the following one. The reason for the difference is
that same-type constraints are checked after type argument deduction.
\begin{lstlisting}
fun foo_2<T>(fun(T)->T f, T a) -> T { return f(a); }

fun id(double x) -> double { return x; }

fun main() -> int@ {
  foo_1(id, 1.0); // ok
  foo_1(id, 1); // error: Same type requirement violated, double != int
  foo_2(id, 1.0); // ok
  foo_2(id, 1); // ok
}
\end{lstlisting}
In the first call to \texttt{foo\_1} the compiler deduces
\texttt{T=double} and \texttt{S=double} from the arguments \texttt{id}
and \texttt{1.0}. The compiler then checks the same-type constraint
\texttt{T == S}, which in this case is satisfied. For the second call
to \texttt{foo\_1}, the compiler deduces \texttt{T=double} and
\texttt{S=int} and then the same-type constraint \texttt{T == S} is
not satisfied. The first call to \texttt{foo\_2} is straightforward.
For the second call to \texttt{foo\_2}, the compiler deduces
\texttt{T=double} from the type of \texttt{id} and the argument
\texttt{1} is implicitly coerced to \texttt{double}.

Type equality is a \Keyword{congruence relation}, which means several
things.  First it means type equality is an \Keyword{equivalence
  relation}, so it is reflexive, transitive, and symmetric. Thus, for
any types $\rho$, $\sigma$, and $\tau$ we have
\begin{itemize}
\item $\tau = \tau$
\item $\sigma = \tau$ implies $\tau = \sigma$
\item $\rho = \sigma$ and $\sigma = \tau$ implies $\rho = \tau$
\end{itemize}
For example, the following function is well formed:
\begin{lstlisting}
fun foo<R,S,T> where { R == S, S == T}
(fun(T)->S f, R a) -> T { return f(a); }
\end{lstlisting}
\noindent The type expression \texttt{R} (the type of \texttt{a})
and the type expression \texttt{T} (the parameter type of \texttt{f})
both denote the same type.

The second aspect of type equality being a congruence is that it
propagates in certain ways with respect to type constructors.  For
example, if we know that \code{S} $=$ \code{T} then we also know that
\code{fun(S)->S} $=$ \code{fun(T)->T}.
Similarly, if we have defined a generic struct such as:
\begin{lstlisting}
struct bar<U> { };  
\end{lstlisting}
then \code{S} $=$ \code{T} implies \code{bar<S>} $=$ \code{bar<T>}.
The propagation of equality also goes in the other direction.
For example, \code{bar<S>} $=$ \code{bar<T>} implies
that \code{S} $=$ \code{T}.
The congruence extends to associated types. So \code{S} $=$ \code{T}
implies \code{C<S>.bar} $=$ \code{C<T>.bar}. However, for associated
types, the propagation does not go in the reverse direction. So
\code{C<S>.bar} $=$ \code{C<T>.bar} does not imply that \code{S} $=$
\code{T}. For example, given the model definitions
\begin{lstlisting}
model C<int> { type bar = bool; };
model C<float> { type bar = bool; };
\end{lstlisting}
we have \code{C<int>.bar} $=$ \code{C<float>.bar} but this
does not imply that \code{int} $=$ \code{float}.

Like type parameters, associated types are in general assumed to be
different from one another. So the following program is ill-formed:
\begin{lstlisting}
concept C<U> { type bar; };
fun foo<S, T> where { C<S>, C<T> } (C<S>.bar a, fun(C<T>.bar)->T f) -> T
{ return f(a); }
\end{lstlisting}
The next program is also ill formed.
\begin{lstlisting}
concept D<U> { type bar; type zow; };
fun foo<T> where { D<T> } (D<T>.bar a, fun(D<T>.zow)->T f) -> T
{ return f(a); }
\end{lstlisting}

In the compiler for \G{} we use the congruence closure algorithm by
Nelson and Oppen~\cite{nelson80:_fast_cong_clos} to keep track of
which types are equal.  The algorithm is efficient: $O(n \log n)$ time
complexity on average, where $n$ is the number of types. It has
$O(n^2)$ time complexity in the worst case. This can be improved by
instead using the Downey-Sethi-Tarjan algorithm which is $O(n \log n)$
in the worst case~\cite{Downey:JACM:1980}.

\subsection{Function overloading and concept-based overloading}
\label{sec:overloading}
\label{sec:concept-based-overloading}

Multiple functions with the same name may be defined and static
overload resolution is performed to decide which function to invoke at
a particular call site. The resolution depends on the argument types
and on the model definitions in scope.  When more than one
overload may be called, the most specific overload is called if one
exists. The basic overload resolution rules are based on those of
\Cpp{}.

In the following simple example, the second \code{foo} is called.

\begin{lstlisting}
fun foo() -> int@ { return -1; }
fun foo(int x) -> int@ { return 0; }
fun foo(double x) -> int@ { return -1; }
fun foo<T>(T x) -> int@ { return -1; }

fun main() -> int@ { return foo(3); }
\end{lstlisting}

\noindent The first \code{foo} has the wrong number of arguments,
so it is immediately dropped from consideration. The second and fourth
are given priority over the third because they can exactly match the
argument type \code{int} (for the fourth, type argument deduction
results in \code{T=int}), whereas the third \code{foo} requires an
implicit coercion from \code{int} to \code{double}.  The second
\code{foo} is favored over the fourth because it is more specific.

A function $f$ is a \Keyword{more specific overload} than function $g$
if $g$ is callable from $f$ but not vice versa.  A function $g$ is
\Keyword{callable from} function $f$ if you could call $g$ from inside
$f$, forwarding all the parameters of $f$ as arguments to $g$, without
causing a type error. More formally, if $f$ has type
$\fun\LT\rep{t_f}\GT \where C_f (\rep{\sigma_f})\arrow\tau_f$ and $g$
has type $\fun\LT\rep{t_g}\GT \where C_g (\rep{\sigma_g})\arrow\tau_g$
then $g$ is callable from $f$ if
\begin{equation*}
  \rep{\sigma_f} \leq [\rep{t_g}/\rep{\rho}]\rep{\sigma_g} 
 \text{ and } C_f \text{ implies } [\rep{t_g}/\rep{\rho}]C_g
\end{equation*}
for some $\rep{\rho}$.

In general there may not be a most specific overload in which case the
program is ill-formed. In the following example, both \code{foo}'s are
callable from each other and therefore neither is more specific.
\begin{lstlisting}
fun foo(double x) -> int@ { return 1; }
fun foo(float x) -> int@ { return -1; }
fun main() -> int@ { return foo(3); }  
\end{lstlisting}
In the next example, neither \code{foo} is callable from
the other so neither is more specific.
\begin{lstlisting}
fun foo<T>(T x, int y) -> int@ { return 1; }
fun foo<T>(int x, T y) -> int@ { return -1; }
fun main() -> int@ { return foo(3, 4); }  
\end{lstlisting}

In Section~\ref{sec:tag-dispatching} we showed how to accomplish
concept-based overloading of several versions of \code{advance} using
the tag dispatching idiom in \Cpp{}. Fig.~\ref{fig:advance-g} shows
three overloads of \code{advance} implemented in \G{}. The signatures
for these overloads are the same except for their \code{where} clauses.
The concept \code{BidirectionalIterator} is a refinement of
\code{InputIterator}, so the second version of \code{advance} is more
specific than the first. The concept \code{RandomAccessIterator} is a
refinement of \code{BidirectionalIterator}, so the third
\code{advance} is more specific than the second.

\begin{fdisplay}{The \code{advance} algorithms using concept-based overloading.}{fig:advance-g}
\begin{lstlisting}
fun advance<Iter> where { InputIterator<Iter> }
(Iter! i, InputIterator<Iter>.difference@ n) {
  for (; n != zero(); --n)
    ++i;
}
fun advance<Iter> where { BidirectionalIterator<Iter> }
(Iter! i, InputIterator<Iter>.difference@ n) {
  if (zero() < n)
    for (; n != zero(); --n)
      ++i;
  else
    for (; n != zero(); ++n)
      --i;
}
fun advance<Iter> where { RandomAccessIterator<Iter> }
(Iter! i, InputIterator<Iter>.difference@ n) {
  i = i + n;
}
\end{lstlisting}
\end{fdisplay}

The code in Fig.~\ref{fig:overload-resolve} shows two calls to
\code{advance}. The first call is with an iterator for a singly-linked
list. This iterator is a model of \code{InputIterator} but not
\code{RandomAccessIterator}; the overload resolution chooses the first
version of \code{advance}.  The second call to \code{advance} is with
a pointer which is a \code{RandomAccessIterator} so the second version
of \code{advance} is called.

\begin{fdisplay}{Example calls to \code{advance} and overload resolution.}{fig:overload-resolve}
  \begin{lstlisting}
use "slist.g";
use "basic_algorithms.g"; // for copy
use "iterator_functions.g"; // for advance
use "iterator_models.g"; // for iterator models for int*

fun main() -> int@ {
  let sl = @slist<int>();
  push_front(1, sl); push_front(2, sl); 
  push_front(3, sl); push_front(4, sl);
  let in_iter = begin(sl);
  advance(in_iter, 2); // calls version 1, linear time

  let rand_iter = new int[4];
  copy(begin(sl), end(sl), rand_iter);
  advance(rand_iter, 2);  // calls version 3, constant time

  if (*in_iter == *rand_iter) return 0;
  else return -1;
}
  \end{lstlisting}
\end{fdisplay}

Concept-based overloading in \G{} is entirely based on static
information available during the type checking and compilation of the
call site. This presents some difficulties when trying to resolve to
optimized versions of an algorithm from within another generic
function. Section~\ref{sec:algo-dispatching} discusses the issues that
arise and presents an idiom that ameliorates the problem.

\subsection{Function expressions}

The following is the syntax for function expressions and function
types.

\noindent The body of a function expression may be either a sequence of
statements enclosed in braces or a single expression following a
colon. The return type of a function expression is deduced from the
return statements in the body, or from the single expression.

The following example computes the sum of an array using
\code{for\_each} and a function expression.
\footnote{Of course, the \code{accumulate} function is
  the appropriate algorithm for this computation, but then the example
  would not demonstrate the use of function expressions.}

\begin{lstlisting}
fun main() -> int@ {
  let n = 8;
  let a = new int[n];
  for (let i = 0; i != n; ++i)
    a[i] = i;
  let sum = 0;
  for_each(a, a + n, fun(int x) p=&sum { *p = *p + x; });
  return sum - (n * (n-1))/2;
}
\end{lstlisting}

\noindent The expression 
\begin{lstlisting}
fun(int x) p=&sum { *p = *p + x; }  
\end{lstlisting}
creates a function object. The body of a function expression is not
lexically scoped, so a direct use of \code{sum} in the body would be
an error. The initialization \code{p=\&sum} declares a data
member inside the function object with type \code{int*} and copy
constructs the member with the address \code{\&sum}.

The primary motivation for non-lexically scoped function expressions
is to keep the design close to \Cpp{} so that function expressions can
be directly compiled to function objects in \Cpp{}. However, this
design has some drawbacks as we discovered while porting the STL to
\G{}.

Most STL implementations implement two separate versions of
\code{find_subsequence}, one written in terms of \code{operator==} and
the other in terms of a function object. The version using
\code{operator==} could be written in terms of the one that takes a
function object, but it is not written that way. The original reason
for this was to improve efficiency, but with with a modern optimizing
compiler there should be no difference in efficiency: all that is
needed to erase the difference is some simple inlining. The \G{}
implementation we write the \code{operator==} version of
\code{find_subsequence} in terms of the higher-order version. The
following code shows how this is done and is a bit more complicated
than we would have liked.

\begin{lstlisting}
fun find_subsequence<Iter1,Iter2>
where { ForwardIterator<Iter1>, ForwardIterator<Iter2>,
        ForwardIterator<Iter1>.value == ForwardIterator<Iter2>.value,
        EqualityComparable<ForwardIterator<Iter1>.value> }
(Iter1 first1, Iter1 last1, Iter2 first2, Iter2 last2) -> Iter1@@
{
  type T = ForwardIterator<Iter1>.value;
  let cmp = model EqualityComparable<T>.operator==;
  return find_subsequence(first1, last1, first2, last2,
                          fun(T a,T b) c=cmp: c(a, b));
}
\end{lstlisting}

\noindent It would have been simpler to write the
function expression as

\begin{lstlisting}
fun(T a, T b): a == b  
\end{lstlisting}

\noindent However, this is an error in \G{} because the \code{operator==}
from the \code{EqualityComparable<..>} requirement is a local name,
not a global one, and is therefore not in scope for the body of the
function expression.  The workaround is to store the comparison
function as a data member of the function object. The expression

\begin{lstlisting}
model EqualityComparable<T>.operator==
\end{lstlisting}

\noindent accesses the \code{operator==} member from the
model of \code{EqualityComparable} for type \code{T}.

Examples such as these are a convincing argument that lexical scoping
should be allowed in function expressions, and the next generation of
\G{} will support this feature.

\subsection{First-class polymorphism}
\label{sec:first-class-poly}

In the introduction we mentioned that \G{} is based on System F. One
of the hallmarks of System F is that it provides first class
polymorphism. That is, polymorphic objects may be passed to and
returned from functions. This is in contrast to the ML family of
languages, where polymorphism is second class. In
Section~\ref{sec:implicit-instantiation} we discussed how the
restriction to second-class polymorphism simplifies type argument
deduction, reducing it to normal unification. However, we prefer to
retain first-class polymorphism and use the somewhat more complicated
variant of unification from \MLF.

One of the reasons to retain first-class polymorphism is to retain the
expressiveness of function objects in \Cpp{}.  A function object may
have member function templates and may therefore by used
polymorphically.
The following program is a simple use of first-class polymorphism in
\G{}. Note that \code{f} is applied to arguments of different types.
\begin{lstlisting}
fun foo(fun<T>(T)->T f) -> int@ { return f(1) + d2i(f(-1.0)); }
fun id<T>(T x) -> T { return x; }
fun main() -> int@ { return foo(id); }
\end{lstlisting}

\section{Analysis of \G{} and the STL}
\label{sec:stl-implementation}

In this section we analyze the interdependence of the language
features of \G{} and generic library design in light of implementing
the STL. A primary goal of generic programming is to express
algorithms with minimal assumptions about data abstractions, so we
first look at how the generic functions of \G{} can be used to
accomplish this. Another goal of generic programming is efficiency, so
we investigate the use of function overloading in \G{} to accomplish
automatic algorithm selection. We conclude this section with a brief
look at implementing generic containers and adaptors in \G{}.

\subsection{Algorithms}

Fig.~\ref{fig:stl-algos-g} depicts a few simple STL algorithms
implemented using generic functions in \G{}.
The STL provides two versions of most algorithms, such as the
overloads for \code{find} in Fig.~\ref{fig:stl-algos-g}.  The first
version is higher-order, taking a predicate function as its third
parameter while the second version relies on \code{operator==}. 
Functions are first-class in \G{}, so the higher-order version is
straightforward to express.
As is typical in the STL, there is a high-degree of internal reuse:
\code{remove} uses \code{remove\_copy} and and \code{find}. 

\begin{fdisplay}{Some STL Algorithms in \G{}.}{fig:stl-algos-g}
\begin{lstlisting}
fun find<Iter> where { InputIterator<Iter> }
(Iter@ first, Iter last, 
 fun(InputIterator<Iter>.value)->bool@ pred) -> Iter@ {
  while (first != last and not pred(*first)) ++first;
  return first;
}
fun find<Iter> where { InputIterator<Iter>,
    EqualityComparable<InputIterator<Iter>.value> }
(Iter@ first, Iter last, InputIterator<Iter>.value value) -> Iter@ {
  while (first != last and not (*first == value)) ++first;
  return first;
}
fun remove<Iter> where { MutableForwardIterator<Iter>,
    EqualityComparable<InputIterator<Iter>.value> }
(Iter@ first, Iter last, InputIterator<Iter>.value value) -> Iter@ {
  first = find(first, last, value);
  let i = @Iter(first);
  return first == last ? first : remove_copy(++i, last, first, value);
}
\end{lstlisting}
\end{fdisplay}

\subsection{Iterators}

Figures \ref{fig:iter-concepts-g-1} and \ref{fig:iter-concepts-g-2}
show the STL iterator hierarchy as represented in \G{}. Required
operations are expressed in terms of function signatures, and
associated types are expressed with a nested \code{type} requirement.
The refinement hierarchy is established with the \code{refines}
clauses and nested model requirements with \code{require}.
The semantic invariants and complexity guarantees of the iterator
concepts are not expressible in \G{} as they are beyond the scope of
its type system.

\begin{fdisplay}{The STL Iterator Concepts in \G{} (Part I).}{fig:iter-concepts-g-1}
\begin{lstlisting}
concept InputIterator<X> {
  type value;
  type difference;
  refines EqualityComparable<X>;
  refines Regular<X>;
  require SignedIntegral<difference>;
  fun operator*(X) -> value@;
  fun operator++(X!) -> X!;
};
concept OutputIterator<X,T> {
  refines Regular<X>;
  fun operator<<(X!, T) -> X!;
};
concept ForwardIterator<X> {
  refines DefaultConstructible<X>;
  refines InputIterator<X>;
  fun operator*(X) -> value;
};
concept MutableForwardIterator<X> {
  refines ForwardIterator<X>;
  refines OutputIterator<X,value>;
  require Regular<value>;
  fun operator*(X) -> value!;
};
\end{lstlisting}
\end{fdisplay}

\begin{fdisplay}{The STL Iterator Concepts in \G{} (Part II).}{fig:iter-concepts-g-2}

\begin{lstlisting}
concept BidirectionalIterator<X> {
  refines ForwardIterator<X>;
  fun operator--(X!) -> X!;
};
concept MutableBidirectionalIterator<X> {
  refines BidirectionalIterator<X>;
  refines MutableForwardIterator<X>;
};
concept RandomAccessIterator<X> {
  refines BidirectionalIterator<X>;
  refines LessThanComparable<X>;
  fun operator+(X, difference) -> X@;
  fun operator-(X, difference) -> X@;
  fun operator-(X, X) -> difference@;
};
concept MutableRandomAccessIterator<X> {
  refines RandomAccessIterator<X>;
  refines MutableBidirectionalIterator<X>;
};
\end{lstlisting}
\end{fdisplay}

\subsection{Automatic Algorithm Selection}
\label{sec:overloading-workaround}

To realize the generic programming efficiency goals, \G{} provides 
mechanisms for automatic algorithm selection.
The following code shows two overloads for \code{copy}. (We omit the
third overload to save space.) The first version is for input
iterators and the second for random access, which uses an integer
counter thereby allowing some compilers to better optimize the loop.
The two signatures are the same except for the \code{where} clause.

\begin{lstlisting}
fun copy<Iter1,Iter2> where { InputIterator<Iter1>,
    OutputIterator<Iter2, InputIterator<Iter1>.value> }
(Iter1@ first, Iter1 last, Iter2@ result) -> Iter2@ {
  for (; first != last; ++first) result << *first;
  return result;
}
fun copy<Iter1,Iter2> where { RandomAccessIterator<Iter1>,
    OutputIterator<Iter2, InputIterator<Iter1>.value> }
(Iter1@ first, Iter1 last, Iter2@ result) -> Iter2@ {
  for (n = last - first; n > zero(); --n, ++first) result << *first;
  return result;
}
\end{lstlisting}

The use of dispatching algorithms such as \code{copy} inside other
generic algorithms is challenging because overload resolution is based
on the surrogate models from the \code{where} clause and not on models
defined for the instantiating type arguments. (This rule is needed for
separate type checking and compilation).  Thus, a call to an
overloaded function such as \code{copy} may resolve to a non-optimal
overload.
Consider the following implementation of
\code{merge}.  The \code{Iter1} and \code{Iter2} types are required to
model \code{InputIterator} and the body of \code{merge} contains two
calls to \code{copy}.

\begin{lstlisting}
fun merge<Iter1,Iter2,Iter3>
where { InputIterator<Iter1>, InputIterator<Iter2>,
        LessThanComparable<InputIterator<Iter1>.value>,
        InputIterator<Iter1>.value == InputIterator<Iter2>.value,
        OutputIterator<Iter3, InputIterator<Iter1>.value> }
(Iter1@ first1, Iter1 last1, Iter2@ first2, Iter2 last2, Iter3@ result)
   -> Iter3@ { ...
  return copy(first2, last2, copy(first1, last1, result));
}  
\end{lstlisting}

\noindent This \code{merge} function always calls the slow version of
\code{copy} even though the actual iterators may be random access.
In \Cpp{}, with tag dispatching, the fast version of \code{copy} is
called because the overload resolution occurs after template
instantiation. However, \Cpp{} does not have separate type checking
for templates.

To enable dispatching for \code{copy}, the type information at the
instantiation of \code{merge} must be carried into the body of
\code{merge} (suppose it is instantiated with a random access
iterator).  This can be done with a combination of concept and model
declarations.  First, define a concept with a single operation that
corresponds to the algorithm.

\begin{lstlisting}
concept CopyRange<I1,I2> {
  fun copy_range(I1,I1,I2) -> I2@;
};
\end{lstlisting}

\noindent Next, add a requirement for this concept to the type requirements
of \code{merge} and replace the calls to \code{copy} with the concept
operation \code{copy\_range}.

\begin{lstlisting}
fun merge<Iter1,Iter2,Iter3>
where { ..., CopyRange<Iter2,Iter3>, CopyRange<Iter1,Iter3> }
(Iter1@ first1, Iter1 last1, Iter2@ first2, Iter2 last2, Iter3@ result)
   -> Iter3@ { ...
  return copy_range(first2, last2, copy_range(first1, last1, result));
}
\end{lstlisting}

\noindent The final step of the idiom is to create parameterized model
declarations for \code{CopyRange}. The \code{where} clauses of the
model definitions match the \code{where} clauses of the respective
overloads for \code{copy}. In the body of each \code{copy\_range} there
is a call to \code{copy} which will resolve to the appropriate overload.

\begin{lstlisting}
model <Iter1,Iter2> where { InputIterator<Iter1>, 
    OutputIterator<Iter2, InputIterator<Iter1>.value> }
CopyRange<Iter1,Iter2> {
  fun copy_range(Iter1 first, Iter1 last, Iter2 result) -> Iter2@
    { return copy(first, last, result); }
};
model <Iter1,Iter2> where { RandomAccessIterator<Iter1>, 
    OutputIterator<Iter2, InputIterator<Iter1>.value> }
CopyRange<Iter1,Iter2> {
  fun copy_range(Iter1 first, Iter1 last, Iter2 result) -> Iter2@
    { return copy(first, last, result); }
};
\end{lstlisting}

A call to \code{merge} with a random access iterator will use the
second model to satisfy the requirement for \code{CopyRange}.  Thus,
when \code{copy\_range} is invoked inside \code{merge}, the fast
version of \code{copy} is called.  A nice property of this idiom is
that calls to generic algorithms need not change. A disadvantage of
this idiom is that the interface of the generic algorithms becomes
more complex.

\subsection{Containers}

The containers of the STL are implemented in \G{} using polymorphic
classes. Fig.~\ref{fig:list} shows an excerpt of the doubly-linked
\code{list} container in \G{}. As usual, a dummy sentinel node is used
in the implementation. With each STL container comes iterator types
that translate between the uniform iterator interface and
data-structure specific operations. Fig.~\ref{fig:list} shows the
\code{list_iterator} which implements \code{operator*} in terms of
\code{x.node->data} and implements \code{operator++} by performing the
assignment \code{x.node = x.node->next}.

\begin{fdisplay}{Excerpt from a doubly-linked list container in \G{}.}{fig:list}
\begin{lstlisting}
struct list_node<T> where { Regular<T>, DefaultConstructible<T> } {
  list_node<T>* next; list_node<T>* prev; T data;
};
class list<T> where { Regular<T>, DefaultConstructible<T> } {
  list() : n(new list_node<T>()) { n->next = n; n->prev = n; }
  ~list() { ... }
  list_node<T>* n;
};
class list_iterator<T> where { Regular<T>, DefaultConstructible<T> } {
  ... list_node<T>* node; 
};
fun operator*<T> where { Regular<T>, DefaultConstructible<T> }
(list_iterator<T> x) -> T { return x.node->data; }

fun operator++<T> where { Regular<T>, DefaultConstructible<T> }
(list_iterator<T>! x) -> list_iterator<T>!
    { x.node = x.node->next; return x; }

fun begin<T> where { Regular<T>, DefaultConstructible<T> }
(list<T> l) -> list_iterator<T>@
    { return @list_iterator<T>(l.n->next); }

fun end<T> where { Regular<T>, DefaultConstructible<T> }
(list<T> l) -> list_iterator<T>@ { return @list_iterator<T>(l.n); }
\end{lstlisting}
\end{fdisplay}

Not shown in Fig.~\ref{fig:list} is the implementation of the
mutable iterator for \code{list} (the \code{list\_iterator} provides
read-only access).  The definitions of the two iterator types are
nearly identical, the only difference is that \code{operator*} returns
by read-only reference for the constant iterator whereas it returns by
read-write reference for the mutable iterator.  The code for these two
iterators should be reused but \G{} does not yet have a language
mechanism for this kind of reuse.

In \Cpp{} this kind of reuse can be expressed using the Curiously
Recurring Template Pattern (CRTP)~\cite{coplien95:_curious} and by
parameterizing the base iterator class on the return type of
\code{operator*}. This approach can not be used in \G{} because the
parameter passing mode may not be parameterized.  Further, the
semantics of polymorphism in \G{} does not match the intended use
here, we want to \emph{generate} code for the two iterator types at
library construction time.  A separate \emph{generative} mechanism is
needed to complement the generic features of \G{}. As a temporary
solution, we used the m4 macro system to factor the common code from
the iterators.  The following is an excerpt from the implementation of
the iterator operators.
\begin{lstlisting}[mathescape=false]
define(`forward_iter_ops',
`fun operator*<T> where { Regular<T>, DefaultConstructible<T> }
($1<T> x) -> T $2 { return x.node->data; } ...')
forward_iter_ops(list_iterator, &) /* read-only */
forward_iter_ops(mutable_list_iter, !) /* read-write */
\end{lstlisting}

\subsection{Adaptors}

The \code{reverse\_iterator} class is a representative example
of an STL adaptor.

\begin{lstlisting}
class reverse_iterator<Iter>
  where { Regular<Iter>, DefaultConstructible<Iter> }
{
  reverse_iterator(Iter base) : curr(base) { }
  reverse_iterator(reverse_iterator<Iter> other) : curr(other.curr) { }
  Iter curr;
};
\end{lstlisting}

\noindent The \code{Regular} requirement on the underlying iterator
is needed for the copy constructor and \code{DefaultConstructible} is
needed for the default constructor.
This adaptor flips the direction of traversal of the underlying
iterator, which is accomplished with the following \code{operator*}
and \code{operator++}.  There is a call to \code{operator--} on the
underlying \code{Iter} type we need to include the requirement for
\concept{Bidirectional Iterator}.

\begin{lstlisting}
fun operator*<Iter> where { BidirectionalIterator<Iter> }
(reverse_iterator<Iter> r) -> BidirectionalIterator<Iter>.value
    { let tmp = @Iter(r.curr); return *--tmp; }

fun operator++<Iter> where { BidirectionalIterator<Iter> }
(reverse_iterator<Iter>! r) -> reverse_iterator<Iter>!
    { --r.curr; return r; }
\end{lstlisting}

\noindent Polymorphic model definitions are used to establish
that \code{reverse\_iterator} is a model of the iterator concepts, as
we discussed in Section~\ref{sec:models}.

\section{The Boost Graph Library}
\label{sec:bgl-implementation}

A group of us at the Open Systems Lab performed a comparative study of
language support for generic
programming~\cite{comparing_generic_programming03}. We evaluated a
half dozen modern programming languages by implementing a subset of
the Boost Graph Library~\cite{siek02:_bgl} in each language. We
implemented a family of algorithms associated with breadth-first
search, including Dijkstra's single-source shortest
paths~\cite{dijkstra59} and Prim's minimum spanning tree
algorithms~\cite{prim57:_short}.  This section extends the previous
study to include \G{}. We give a brief overview of the BGL, describe
the implementation of the BGL in \G{}, and compare the results to
those in our earlier study~\cite{comparing_generic_programming03}.

\subsection{An overview of the BGL graph search algorithms}

\begin{figure*}[hbtp]
 \centerline{\includegraphics[width=5in]{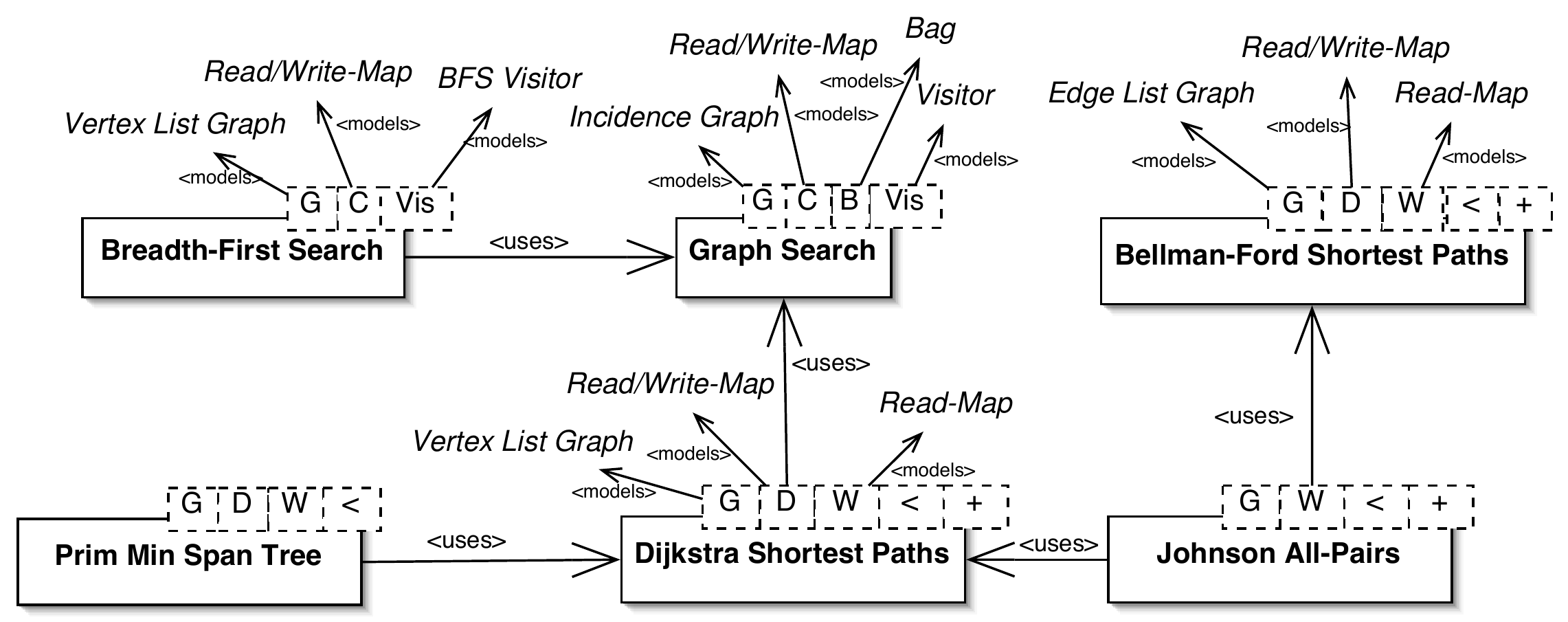}}
 \caption{Graph algorithm parameterization and reuse within the Boost Graph Library. 
   Arrows for redundant models relationships are not shown.  For
   example, the type parameter \code{G} of breadth-first search
   must also model \cpt{Incidence Graph} because breadth-first search uses
   graph search.}
 \label{fig:algo-param}
\end{figure*}

Figure~\ref{fig:algo-param} depicts some graph search algorithms from
the BGL, their relationships, and how they are parameterized. Each
large box represents an algorithm and the attached small boxes
represent type parameters.  An arrow labeled \code{<uses>} from one
algorithm to another specifies that one algorithm is implemented using
the other.  An arrow labeled \code{<models>} from a type parameter to
an unboxed name specifies that the type parameter must model that
concept.  For example, the breadth-first search algorithm has
three type parameters: \code{G}, \code{C}, and \code{Vis}.  Each of
these has requirements: \code{G} must model the \cpt{Vertex List
  Graph} and \cpt{Incidence Graph} concepts, \code{C} must model the
\cpt{Read/Write Map} concept, and \code{Vis} must model the \cpt{BFS
  Visitor} concept. The breadth-first search algorithm is
implemented using the graph search algorithm.

The core algorithm of this library is graph search, 
which traverses a graph and performs 
user-defined operations at certain points in the search. 
The order in which vertices are visited is controlled by a type
argument, \code{B}, that models the \cpt{Bag} concept. This concept
abstracts a data structure with insert and remove operations but no
requirements on the order in which items are removed.  When \code{B}
is bound to a FIFO queue, the traversal order is breadth-first.  When
it is bound to a priority queue based on distance to a source vertex,
the order is closest-first, as in Dijkstra's single-source shortest
paths algorithm.  Graph search is also parameterized on actions
to take at event points during the search, such as when a vertex is
first discovered.  This parameter, \code{Vis}, must model the
\cpt{Visitor} concept (which is not to be confused with the Visitor
design pattern).  The graph search algorithm also takes a type
parameter \code{C} for mapping each vertex to a color and \code{C}
must model the \cpt{Read/Write Map} concept. The colors are used as
markers to keep track of the progression of the algorithm through the
graph.

The \cpt{Read Map} and \cpt{Read/Write Map} concepts represent
variants of an important abstraction in the graph library: the
\Keyword{property map}.  In practice, graphs represent domain-specific
entities. For example, a graph might depict the layout of a
communication network, its vertices representing endpoints and its
edges representing direct links.  In addition to the number of
vertices and the edges between them, a graph may associate values to
its elements.  Each vertex of a communication network graph might have
a name and each edge a maximum transmission rate.  Some algorithms
require access to domain information associated with the graph
representation.  For example, Prim's minimum spanning tree
algorithm requires ``weight'' information associated with each edge in
a graph.  Property maps provide a convenient implementation-agnostic
means of expressing, to algorithms, relations between graph elements
and domain-specific data. Some graph data structures directly contain
associated values with each node; others use external associative data
structures to implement these relationships.  Interfaces based on
property maps work equally well with both representations.

The graph algorithms are all parameterized on the graph type.
Breadth-first search takes a type parameter \code{G}, which must
model two concepts, \cpt{Incidence Graph} and \cpt{Vertex List
  Graph}.  The \cpt{Incidence Graph} concept defines an interface
for accessing out-edges of a vertex.  \cpt{Vertex List Graph}
specifies an interface for accessing the vertices of a graph in an
unspecified order.  The Bellman-Ford shortest paths
algorithm~\cite{bellman58} requires a model of the \cpt{Edge List
  Graph} concept, which provides access to all the edges of a graph.

That graph capabilities are partitioned among three concepts
illustrates generic programming's emphasis on minimal algorithm requirements.
The Bellman-Ford shortest paths algorithm requires of a graph
only the operations described by the \cpt{Edge List Graph}
concept.  Breadth-first search, in contrast, requires the functionality
of two separate concepts.  By partitioning the
functionality of graphs, each algorithm
can be used with any data type that meets its minimum requirements.
If the three fine-grained graph concepts were replaced with
one monolithic concept, each algorithm
would require more from its graph type parameter than 
necessary and would thus unnecessarily restrict the set of types
with which it could be used.

The graph library design is suitable for evaluating generic
programming capabilities of languages because its implementation
involves a rich variety of generic programming techniques.  Most of
the algorithms are implemented using other library algorithms:
breadth-first search and Dijkstra's shortest paths use
graph search, Prim's minimum spanning tree algorithm
uses Dijkstra's algorithm, and Johnson's all-pairs
  shortest paths algorithm~\cite{johnson77:_short_paths} uses both
Dijkstra's and Bellman-Ford shortest paths.
Furthermore, type parameters for some algorithms, such as the \code{G}
parameter to breadth-first search, must model multiple
concepts.  In addition, the algorithms require certain relationships
between type parameters. For example, consider the graph search
algorithm.  The \code{C} type argument, as a model of \cpt{Read/Write
  Map}, is required to have an associated key type.  The \code{G} type
argument is required to have an associated vertex type.  Graph
  search requires that these two types be the same.

As in our earlier study, we focus the evaluation on the interface of
the breadth-first search algorithm and the infrastructure
surrounding it, including concept definitions and an example use of
the algorithm.

\subsection{Implementation in \G{}}

So far we have implemented breadth-first search and Dijkstra's
single-source shortest paths in \G{}. This required defining several
of the graph and property map concepts and an implementation of the
\texttt{adjacency\_list} class, a FIFO queue, and a priority queue.
 
The interface for the breadth-first search algorithm is
straightforward to express in \G{}.  It has three type parameters: the
graph type \texttt{G}, the color map type \texttt{C}, and the visitor
type \texttt{Vis}. The requirements on the type parameters are
expressed with a \texttt{where} clause, using concepts that we
describe below. In the interface of \texttt{breadth\_first\_search},
associated types and same-type constraints play an important role in
accurately tracking the relationships between the graph type, its
vertex descriptor type, and the color property map.

\begin{lstlisting}
type Color = int;
let black = 0;
let gray  = 1;
let white = 2;

fun breadth_first_search<G, C, Vis>
  where { IncidenceGraph<G>, VertexListGraph<G>,
          ReadWritePropertyMap<C>,
          PropertyMap<C>.key == IncidenceGraph<G>.vertex_descriptor,
          PropertyMap<C>.value == Color,
          BFSVisitor<Vis,G> }
(G g, IncidenceGraph<G>.vertex_descriptor@ s, C c, Vis vis) { /* ... */ }
\end{lstlisting}

Figure~\ref{fig:graph-concepts} shows the definition of several graph
concepts in \G{}. The \texttt{Graph} concept requires the associated
types \texttt{vertex\_descriptor} and \texttt{edge\_descriptor} and
some basic functionality for those types such as copy construction and
equality comparison. This concept also includes the \texttt{source}
and \texttt{target} functions. The \texttt{Graph} concept serves to
factor common requirements out of the \texttt{IncidenceGraph} and
\texttt{VertexListGraph} concepts.

The \texttt{IncidenceGraph} concept introduces the capability to
access out-edges of a vertex.  The access is provided by the
\texttt{out\_edge\_iterator} associated type. The requirements for the
out-edge iterator are slightly more than the standard
\texttt{InputIterator} concept and slightly less than the
\texttt{ForwardIterator} concept. The out-edge iterator must allow for
multiple passes but dereferencing an out-edge iterator need not return
a reference (for example, it may return by-value instead). Thus we
define the following new concept to express these requirements.

\begin{lstlisting}
concept MultiPassIterator<Iter> {
    refines DefaultConstructible<Iter>;
    refines InputIterator<Iter>;
    // semantic requirement: allow multiple passes through the range
};
\end{lstlisting}

\noindent In Figure~\ref{fig:graph-concepts}, the \texttt{IncidenceGraph} concept
uses same-type constraints\index{same-type constraints} to require
that the \texttt{value} type of the iterator to be the same type as
the \texttt{edge\_descriptor}.
The \texttt{VertexListGraph} concepts adds the capability of
traversing all the vertices in the graph using the associated
\texttt{vertex\_iterator}.

\begin{fdisplay}{Graph concepts in \G{}.}{fig:graph-concepts}
\begin{lstlisting}
concept Graph<G> {
    type vertex_descriptor;
    require DefaultConstructible<vertex_descriptor>;
    require Regular<vertex_descriptor>; 
    require EqualityComparable<vertex_descriptor>;

    type edge_descriptor;
    require DefaultConstructible<edge_descriptor>;
    require Regular<edge_descriptor>;
    require EqualityComparable<edge_descriptor>;

    fun source(edge_descriptor, G) -> vertex_descriptor@;
    fun target(edge_descriptor, G) -> vertex_descriptor@;
};
concept IncidenceGraph<G> {
    refines Graph<G>;

    type out_edge_iterator;
    require MultiPassIterator<out_edge_iterator>;
    edge_descriptor == InputIterator<out_edge_iterator>.value;

    fun out_edges(vertex_descriptor, G)
        -> pair<out_edge_iterator, out_edge_iterator>@;
    fun out_degree(vertex_descriptor, G) -> int@;
};
concept VertexListGraph<G> { 
    refines Graph<G>;
    
    type vertex_iterator;
    require MultiPassIterator<vertex_iterator>;
    vertex_descriptor == InputIterator<vertex_iterator>.value;

    fun vertices(G) -> pair<vertex_iterator, vertex_iterator>@;
    fun num_vertices(G) -> int@;
};
\end{lstlisting}
\end{fdisplay}

Figure~\ref{fig:vector-as-graph} shows the implementation of a graph
in terms of a vector of singly-linked lists. Vertex descriptors are
integers and edge descriptors are pairs of integers. The out-edge
iterator is implemented with the \texttt{vg\_out\_edge\_iter} class
whose implementation is shown in Figure~\ref{fig:vg-out-edge-iter}.
The basic idea behind this iterator is to provide a different view of
the list of target vertices, making it appear as a list of
source-target pairs.

\begin{fdisplay}{Implementation of a graph with a vector of lists.}{fig:vector-as-graph}
\begin{lstlisting}
fun source(pair<int,int> e, vector< slist<int> >) -> int@ 
  { return e.first; }
fun target(pair<int,int> e, vector< slist<int> >) -> int@ 
  { return e.second; }

model Graph< vector< slist<int> > > {
  type vertex_descriptor = int;
  type edge_descriptor = pair<int,int>;
};

fun out_edges(int src, vector< slist<int> > G)
    -> pair<vg_out_edge_iter, vg_out_edge_iter>@ { 
  return make_pair(@vg_out_edge_iter(src, begin(G[src])),
                   @vg_out_edge_iter(src, end(G[src])));
}
fun out_degree(int src, vector< slist<int> > G) -> int@ 
  { return size(G[src]); }

model IncidenceGraph< vector< slist<int> > > {
  type out_edge_iterator = vg_out_edge_iter;
};

fun vertices(vector< slist<int> > G) 
  -> pair<counting_iter,counting_iter>@
  { return make_pair(@counting_iter(0), @counting_iter(size(G))); }
fun num_vertices(vector< slist<int> > G) -> int@ { return size(G); }

model VertexListGraph< vector< slist<int> > > {
  type vertices_size_type = int;
  type vertex_iterator = counting_iter;
};
\end{lstlisting}
\end{fdisplay}

\begin{fdisplay}{Out-edge iterator for the vector of lists.}{fig:vg-out-edge-iter}
\begin{lstlisting}
class vg_out_edge_iter {
  vg_out_edge_iter() { }
  vg_out_edge_iter(int src, slist_iterator<int> iter) 
    : src(src), iter(iter) { }
  vg_out_edge_iter(vg_out_edge_iter x) 
    : iter(x.iter), src(x.src) { }
  slist_iterator<int> iter;
  int src;
};
fun operator=(vg_out_edge_iter! me, vg_out_edge_iter other) 
  -> vg_out_edge_iter!
  { me.iter = other.iter; me.src = other.src; return me; }
model DefaultConstructible<vg_out_edge_iter> { };
model Regular<vg_out_edge_iter> { };

fun operator==(vg_out_edge_iter x, vg_out_edge_iter y) -> bool@
  { return x.iter == y.iter; }
fun operator!=(vg_out_edge_iter x, vg_out_edge_iter y) -> bool@
  { return x.iter != y.iter; }
model EqualityComparable<vg_out_edge_iter> { };

fun operator*(vg_out_edge_iter x) -> pair<int,int>@
  { return make_pair(x.src, *x.iter); }
fun operator++(vg_out_edge_iter! x) -> vg_out_edge_iter!
  { ++x.iter; return x; }
model InputIterator<vg_out_edge_iter> {
  type value = pair<int,int>;
  type difference = ptrdiff_t;
};
model MultiPassIterator<vg_out_edge_iter> { };
\end{lstlisting}
\end{fdisplay}

The property map concepts are defined in
Figure~\ref{fig:property-maps}. The \texttt{ReadWritePropertyMap} is a
refinement of the \texttt{ReadablePropertyMap} concept, which requires
the \texttt{get} function, and the \texttt{WritablePropertyMap}
concept, which requires the \texttt{put} function.  Both of these
concepts refine the \texttt{PropertyMap} concept which includes the
associated \texttt{key} and \texttt{value} types.

\begin{fdisplay}{Property map concepts in \G{}.}{fig:property-maps}
\begin{lstlisting}
concept PropertyMap<Map> {
    type key;
    type value;
};
concept ReadablePropertyMap<Map> {
    refines PropertyMap<Map>;
    fun get(Map, key) -> value;
};
concept WritablePropertyMap<Map> {
    refines PropertyMap<Map>;
    fun put(Map, key, value);
};
concept ReadWritePropertyMap<Map>  {
    refines ReadablePropertyMap<Map>;
    refines WritablePropertyMap<Map>;
};
\end{lstlisting}
\end{fdisplay}

Figure~\ref{fig:bfs-visitor} shows the definition of the
\texttt{BFSVisitor} concept. This concept is naturally expressed as a
multi-parameter concept because the visitor and graph types are
independent: a particular visitor may be used with many different
concrete graph types and vice versa.
The use of \texttt{refines} for \texttt{Graph} in \texttt{BFSVisitor}
is somewhat odd, \texttt{require} would be more natural, but the
refinement provides direct (and convenient) access to the vertex and
edge descriptor types. An alternative would be use to \texttt{require}
and some type aliases, but type aliases have not yet been added to
concept definitions.

\begin{fdisplay}{Breadth-first search visitor concept.}{fig:bfs-visitor}
\begin{lstlisting}
concept BFSVisitor<Vis, G> {
    refines Regular<Vis>;
    refines Graph<G>;

    fun initialize_vertex(Vis v, vertex_descriptor d, G g) {}
    fun discover_vertex(Vis v, vertex_descriptor d, G g) {}
    fun examine_vertex(Vis v, vertex_descriptor d, G g) {}
    fun examine_edge(Vis v, edge_descriptor d, G g) {}
    fun tree_edge(Vis v, edge_descriptor d, G g) {}
    fun non_tree_edge(Vis v, edge_descriptor d, G g) {}
    fun gray_target(Vis v, edge_descriptor d, G g) {}
    fun black_target(Vis v, edge_descriptor d, G g) {}
    fun finish_vertex(Vis v, vertex_descriptor d, G g) {}
};
\end{lstlisting}
\end{fdisplay}

Figure~\ref{fig:bfs-example} presents an example use of the
\code{breadth_first_search} function to output vertices in
breadth-first order.  To do so, the \code{test_vis} visitor overrides
the function \code{discover_vertex}; empty implementations of the
other visitor functions are provided by \code{default_bfs_visitor}. A
graph is constructed using the \code{AdjacencyList} class, and then
\code{breadth_first_search} is called.

\begin{fdisplay}{Example use of the BFS generic function.}{fig:bfs-example}
\begin{lstlisting}
struct test_vis { };
fun discover_vertex<G>(test_vis, int v, G g) { printf("%d ", v); }

model <G> where { Graph<G>,  Graph<G>.vertex_descriptor == int } 
BFSVisitor<test_vis, G> { };

fun main() -> int@ {
  let n = 7;
  let g = @vector< slist<int> >(n);
  push_front(1, g[0]); push_front(4, g[0]);
  push_front(2, g[1]); push_front(3, g[1]);
  push_front(4, g[3]); push_front(6, g[3]);
  push_front(5, g[4]);

  let src = 0;
  let color = new Color[n];
  for (let i = 0; i != n; ++i)
    color[i] = white;
  breadth_first_search(g, src, color, @test_vis());
  return 0;
}
\end{lstlisting}
\end{fdisplay}

\section{Related Work}
\label{sec:related-work}

There is a long history of programming language support for
polymorphism, dating back to the
1970s~\cite{GIRARD72,REYNOLDS74C,Liskov77:CLU,Milner78}. An early
precursor to \G{}'s concept feature can be seen in CLU's type set
feature~\cite{Liskov77:CLU}. Type sets differ from concepts in that
they rely on structural conformance whereas concepts use nominal
conformance established by a \code{model} definition. Also, \G{}
provides a means for composing concepts via refinement whereas CLU
does not provide a means for composing type sets. Finally, CLU does
not provide support for associated types.  

In mathematics, the notion of algebraic structure is equivalent to
\G{}'s concept, and has been in use for a very long
time~\cite{bourbaki68:_theory_sets}.

\textbf{Type classes}
The concept feature in \G{} is heavily influenced by the type class
feature of Haskell~\cite{Wadler89}, with its nominal conformance and
explicit model definitions. However, \G{}'s support for associated
types, same type constraints, and concept-based overloading is novel.
Also, \G{}'s type system is fundamentally different from Haskell's: it
is based on System F~\cite{GIRARD72,REYNOLDS74C} instead of
Hindley-Milner type inference~\cite{Milner78}. This difference has
some repercussions.  In \G{} there is more control over the scope of
concept operations because \code{where} clauses introduce concept
operations into the scope of the body. This difference allows Haskell
to infer type requirements but induces the restriction that two type
classes in the same module may not have operations with the same name.
A difference we discussed in Section~\ref{sec:scoped-models} is that
in \G{}, overlapping models may coexist in separate scopes but still
be used in the same program, whereas in Haskell overlapping models may
not be used in the same program.  Haskell performed quite well in our
comparative study of support for generic
programming~\cite{comparing_generic_programming03}. However, we
pointed out that Haskell was missing support for associated types and
work to remedy this has been reported
in~\cite{chakravarty04:_assoc_types,chakravarty05:_assoc_type_syn}.

Wehr, Lammel, and Thiemann\cite{Wehr:2007uq} have proposed extending
Java with generalized interfaces, which bear a close resemblance to
\G{}'s concepts and Haskell's type classes, but add the capability of
run-time dispatch using existential quantification. (\G{} currently
provides only universal quantification, although programmers can
workaround this limitation with an tricky
encoding~\cite{Pierce:2002hj}).

\textbf{Signatures and functors}
A rough analogy can be made between SML
signatures~\cite{milner90:definition_sml} and \G{} concepts, and
between ML structures and \G{} models.  However, there are significant
differences.  Functors are module-level constructs and therefore
provide a more coarse-grained mechanism for parameterization than do
generic functions. More importantly, functors require explicit
instantiation with a structure, thereby making their use more
heavyweight than generic functions in \FG{}, which perform automatic
lookup of the required model or instance. The associated types and
same-type constraints of \G{} are roughly equivalent to types nested
in ML signatures and to type sharing respectively.  We reuse some
implementation techniques from ML such as a union/find-based algorithm
for deciding type equality~\cite{macqueen88implementation}.
There are numerous other languages with parameterized
modules~\cite{poll99:_aldor,Goguen:OBJ,taft97:_ada_manual} that 
require explicit instantiation with a structure.

Recently, Dreyer, Harper, Chakravarty, and Keller proposed an
extension of SML signatures/functors, call modular type
classes~\cite{Dreyer:2007fk}, that provides many of the benefits of
Haskell type classes such as implicit instantiation and instance
passing. The design for modular type classes differs from concepts in
\G{} primarily in that it supports the convenience of type inference
at the price of disallowing overlapping instances in a given scope and
first-class polymorphism.

\textbf{Subtype-bound polymorphism}
Less closely related to \G{} are languages based on subtype-bounded
polymorphism~\cite{canning89:_f_bound_poly} such as Java, C\#, and
Eiffel. We found subtype-bounded polymorphism less suitable for
generic programming and refer the reader to
\cite{comparing_generic_programming03} for an in-depth discussion.

\textbf{Row variable polymorphism}
OCaml's object types\cite{leroy03:_ocaml,Remy-Vouillon!tapos} and
polymorphism over row variables provide fairly good support for
generic programming. However, OCaml lacks support for associated
types so it suffers from clutter due to extra type parameters in
generic functions.  PolyTOIL~\cite{bruce95polytoil}, with its
match-bound polymorphism, provides similar support for generic
programming as OCaml but also lacks associated types.

\textbf{Virtual types}
One of the proposed solutions for dealing with binary methods and
associated types in object-oriented languages is \Keyword{virtual
  types}, that is, the nesting of abstract types in interfaces and
type definitions within classes or objects. The beginning of this line
of research was the \Keyword{virtual patterns} feature of the
BETA\index{BETA} language~\cite{kristensen83:_beta}. Patterns are a
generalization of classes, objects, and procedures. An adaptation of
virtual patterns to object-oriented classes, called \Keyword{virtual
  classes}, was created by Madsen and
Moller-Pedersen~\cite{madsen89:_virtual_classes} and an adaptation for
Java was created by Thorup~\cite{Thorup97}.  These early designs for
virtual types were not statically type safe, but relied on dynamic
type checking. However, a statically type safe version was created by
Torgersen~\cite{torgersen98:_virtual_types_safe}.  A statically type
safe version of BETA's virtual patterns was developed for the
gbeta\index{gbeta} language of
Ernst~\cite{ernst99b,ernst01:_famil_polym}; the Scala\index{Scala}
programming language also includes type safe virtual types
\cite{odersky-et-al:ecoop03,scala-overview-tech-report}.

\section{Conclusion}
\label{sec:conclusion}

This article presents a new programming language named \G{} that is
designed to meet the needs of large-scale generic libraries.  We
demonstrated this with an implementation of the Standard Template
Library (STL) and the Boost Graph Library (BGL).
We were able to implement all of the abstractions in the STL and BGL
in a straightforward manner.
Further, \G{} is particularly well-suited for the development of
reusable components due to its support of modular type checking and
separate compilation.
\G{}'s strong type system provides support for the independent
validation of components and \G{}'s system of concepts and constraints
allows for rich interactions between components without sacrificing
encapsulation.
The language features present in \G{} promise to increase programmer
productivity with respect to the development and use of generic
components.

\section*{Acknowledgments}

We thank Ronald Garcia, Jeremiah Willcock, Doug Gregor, Jaakko
J\"arvi, Dave Abrahams, Dave Musser, and Alexander Stepanov for many
discussions and collaborations that informed this work.  This work was
supported by NSF grants EIA-0131354 and CCF-0702362, and by a grant
from the Lilly Endowment.

\bibliography{local,eiffel,csharp,implicit-inst,comp_gp_refs,generic-programming,c++_standards_proposals,lsconly,cad,linear-algebra,ggcl,optimization,agse-biblio,formal-specification}
\bibliographystyle{splncs}

\end{document}